\documentclass[adp,fleqn]{w-art}
\usepackage{times}
\usepackage{w-thm}
\usepackage[]{graphicx}
\chardef\bslash=`\\ % p. 424, TeXbook

\usepackage{amsfonts}
\usepackage{amsmath}
\usepackage{amssymb}
\usepackage{psfrag}
\usepackage{graphicx}%
\newcommand{\be}{\begin{equation}}
\newcommand{\ee}{\end{equation}}
\newcommand{\bea}{\begin{eqnarray}}
\newcommand{\eea}{\end{eqnarray}}

\begin{document}
%%
%%  Most of the following commands will be completed by the publisher.
%%
%%  The copyrightyear is defined in the .clo file as the first argument
%%  of the copyrightinfo command. If the copyrightyear differs from that
%%  value it might be adjusted by the following definition:
%%
%% \renewcommand{\copyrightyear}{2002}% uncomment to change the copyrightyear.
%%
\DOIsuffix{theDOIsuffix}
%%
%% issueinfo for header and copyright line
%%
\Volume{12}
\Issue{1}
\Copyrightissue{01}
\Month{01}
\Year{2003}
%%
%%  First and last pagenumber of the article. If the option
%%  'autolastpage' is set (default) the second argument may be left empty.
\pagespan{1}{}
%%
%%    Dates will be filled in by the publisher. The 'reviseddate' and
%%    'dateposted' (Published online) entry may be left empty.
\Receiveddate{29 September 2009}
\Reviseddate{13 November 2009}
\Accepteddate{19 November 2009 by U. Eckern}
%% \Dateposted{3 December 1900}
%%
%%
 % up to three, separated by commas

%72.15.Qm : Scattering mechanisms and Kondo effect 
%           (see also 75.20.Hr Local moments in compounds 
%            and alloys; Kondo effect, valence fluctuations, 
%            heavy fermions in magnetic properties and materials)

%73.63.Kv 	Quantum dots 

%% \pretitle{Editor's Choice}

%% We have a short and a long form for the title. The short form
%% (optional argument) goes into the running head.

\title[Transport through a Kondo quantum dot]{Transport through a Kondo quantum dot: Functional RG \\approach}

%% Please do not enter footnotes or \inst{}-notes into the optional
%% argument of the author command. The optional argument will go into
%% the header. If there is only one address the marker \inst{x} may be
%% omitted.

%%   Information for the first author.
\author[Schmidt]{H. Schmidt \inst{1}} 
\address[\inst{1}]{Institut f\"{u}r Theorie der Kondensierten Materie and DFG-Center for
Functional Nanostructures, Universit\"{a}t Karlsruhe, 76128 Karlsruhe, Germany}
\address[\inst{2}]{Institut f\"{u}r Nanotechnologie, Forschungszentrum Karlsruhe, 76021 Karlsruhe, Germany}
%%
%%    Information for the second author
\author[Woelfle]{P. W\"olfle\footnote{Corresponding
     author \quad E-mail: {\sf woelfle@tkm.uni-karlsruhe.de}, Phone: +49\,721\,608\,3590}\inst{1,2}}

%%
%%    Information for the third author
%%\author[Th.\ Author]{Third Author\footnote{Optional, e. g. e-mail}\inst{2}}
%%
%%  \dedicatory{This is a dedicatory.}

%%
\begin{abstract}
 We apply the functional renormalization group (FRG) method to calculate the
conductance of a quantum dot in the Kondo regime. Starting from the exact FRG
equations in Keldysh formulation for the Kondo exchange Hamiltonian in
pseudo-fermion (pf) representation, we solve the coupled equations for the pf
self energy and the coupling function, neglecting three-particle and higher
correlation functions. The conductance $G$ as a function of temperature $T$
and bias voltage $V$ is calculated using a renormalized Golden Rule
expression. The limiting behavior at $T$ and/or $V$ $\gg T_{K}$ ($T_{K}$ :
Kondo temperature) agrees with known results. The difficulties when
approaching strong coupling are analyzed and improvements are suggested.
\end{abstract}

\maketitle

\section{Introduction}

The Kondo effect is a ubiquitous phenomenon in the area of electron
transport in nanostructures \cite{Wiel, Gold, Cron, Wiel2, Nygard, Jesp}.
For its existence a small system with internal degrees of freedom (e.g. a
quantum dot with spin), coupled to a sufficiently large conduction electron
system (the leads) is required. If certain symmetries are satisfied, the
localized spin forms a resonance state with the surrounding conduction
electron spins at the Fermi level, of energy width $T_{K}$ , the Kondo
temperature (here and in the following we use units where Boltzmann's
constant $k_{B}=1$ and Planck's constant $\hbar =1$ ) . \ A finite bias
voltage $V$ applied across the quantum dot is known to destroy the Kondo
resonance, provided $eV\gg T_{K}$. It has been argued that the limitation in
the coherence time brought about by inelastic processes in the leads causes
the suppression of the Kondo effect at finite $V.$, since it cuts off the
renormalization group flow of the coupling constant \cite{Kam, Rosch}. At
sufficiently large $V$ this cutoff is still within the perturbative regime.
In that case the perturbative RG treatment ("poor man's scaling", \cite{And}
)\ generalized to nonequilibrium, proposed in \cite{pertRG1}, \ provides a
controlled and complete description of all physical quantities. Indeed the
latter theory has been used to decribe experimental data for various
situations, each time with excellent success \cite{pertRG2, Paaske3, Koert1,
Koert2}. While the effect of decoherence on the RG flow of the coupling
constant has been included in \cite{pertRG1,pertRG2} in a phenomenological
way, it is of interest to derive the effect of the decoherence rate in a
systematic fashion. The way in which the decoherence rate cuts off the
logarithmic singularities in perturbation theory has been analyzed in \cite%
{Paaske2}, where it has been found that the decoherence appears in the form
of physically accessible rates, namely the longitudinal and transverse spin
relaxation rates. In a completely new way the effect of decoherence on the
Kondo effect has been analyzed within the flow equation method \cite%
{Kehrein1}, with results in agreement with and thus confirming the work by
Paaske et al. \cite{pertRG1,Paaske2}. The latter method is, however, not
easily applicable to the calculation of observable quantities like the
conductance. It is therefore worthwhile to work out a systematic and
consistent derivation of the coupled RG-equations for the coupling function
and the decoherence rate, even in the weak coupling regime. 

On the other hand, it has not been possible , so far, to calculate the
influence of a finite bias voltage in the strong coupling regime. There are
several methods available describing properties of the Kondo system at
strong coupling in equilibrium (see \cite{Hewson}), but a general and
practically applicable extension of these methods to nonequilibrium has not
been found yet. A recently proposed \ "Conserving T-matrix approximation ",
based on self-consistent renormalized perturbation theory including all
two-particle channels, which has been shown to recover correctly the Fermi
liquid behavior at low energies and the crossover to the local moment regime 
\cite{Kirch, Kroha}, may in principle be generalized to nonequilibrium, but
is at present too costly in numerical evaluation effort.

In this paper we present results for the weak-coupling regime of the Kondo
model in and out of equilibrium obtained by the functional renormalization
group (FRG) method \cite{Wett, Salm, Med, Metz}. The FRG is a powerful tool
when dealing with the diversity of energy scales and the interplay of
correlations in an interacting fermionic systems such as the Kondo model
where the microscopic energy scale band width $D$ and the dynamically
generated energy scale Kondo temperature $T_{K}$ differ by orders of
magnitude. The FRG uses an exact mapping of the initial model to models with
reduced band width. The mapping generates an infinite hierarchy of flow
equations, describing the gradual evolution from a microscopic model
Hamiltonian to the effective action as a function of a continuously
decreasing energy cutoff. In practice, one modifies the bare propagator by
cutting off frequencies smaller then some infrared cutoff $\Lambda$. As this
cutoff is progressively lowered more and more low-energy degrees of freedom
are included, until the original model is recovered for $\Lambda\rightarrow0$%
. The conventionally used approximation consists of a suitable truncation of
the hierarchy of equations. We find that the truncation neglecting three
particle and higher order correlation functions, and keeping only the self
energy together with the four-point vertex function captures the weak
coupling regime ($T,V\gg T_{K}$) completely. In this paper we show in detail
how the transport properties of a Kondo quantum dot in the weak-coupling
regime may be calculated in a systematic and complete way.

Compared to previous applications of the FRG method to Kondo physics (\cite%
{Jak} and ref. therein) our approach differs in at least two ways: (1) the
present work addresses the Kondo model, rather than the Anderson model
(compare, however, the recent work \cite{Schuricht} in which the dynamical
spin response of the Kondo model has been considered, using a real-time RG
method \cite{Schoeller}); (2) in contrast to previous FRG works we include
the effect of the decoherence rate.

\section{Model and FRG formalism}

We consider the model of a quantum dot described by a spin $\frac{1}{2}$
operator $\mathbf{S}$ coupled by exchange interaction and tunneling to the
conduction electron spins as described by the Hamiltonian 
\begin{equation*}
H={\sum\limits_{k\alpha\sigma}}(\epsilon_{k}-\mu_{\alpha})c_{k\alpha\sigma
}^{\dagger}c_{k\alpha\sigma}^{\phantom{\dagger}}+{J\sum\limits_{\alpha
,\alpha^{\prime}=1,2}}\ \mathbf{S}\cdot\mathbf{s}_{\alpha\alpha^{\prime}}
\end{equation*}
Here $\alpha,\alpha^{\prime}=1,2$ labels the leads assumed to be in
equilibrium at chemical potentials $\mu_{1,2}=\pm eV/2$ and $\sigma
=\uparrow,\downarrow$ are spin indices. The conduction electron spin density
operators at the quantum dot are defined as \ $\mathbf{s}_{\alpha
\alpha^{\prime}}={\textstyle\sum\limits_{k,k^{\prime},\sigma,\sigma^{\prime}}%
}\frac{1}{2}c_{k\alpha\sigma}^{\dagger}\mathbf{\tau}_{\sigma,\sigma\prime
}c_{k^{\prime}\alpha\prime\sigma^{\prime}}^{\phantom{\dagger}}$\ , with $%
c_{k\alpha\sigma}^{\dagger}$ an electron creation operator in momentum state 
$k$, lead $\alpha$ and spin state $\sigma$ and $\mathbf{\tau}$ the vector of
Pauli matrices. The $\mathbf{s}_{\alpha\alpha\prime}$ are tunneling
operators for $\alpha\neq\alpha^{\prime}$. We use a pseudo-fermion
representation for the local spin $\mathbf{S=}\frac{1}{2}\sum
\limits_{\gamma,\gamma\prime}f_{\gamma}^{\dagger}\mathbf{\tau}%
_{\gamma\gamma\prime}f_{\gamma\prime }^{\phantom{\dagger}}$ , with
projection onto the physical Hilbert space by adding a term $%
\lambda\lbrack\sum \limits_{\alpha}f_{\gamma}^{\dagger}f_{\gamma}^{%
\phantom{\dagger}}-1]$ to the Hamiltonian and taking the limit $%
\lambda\rightarrow\infty$ (cf. \cite{Abri,Paaske1}).

\begin{figure}[t]
\centering
\includegraphics[width=0.98\columnwidth]{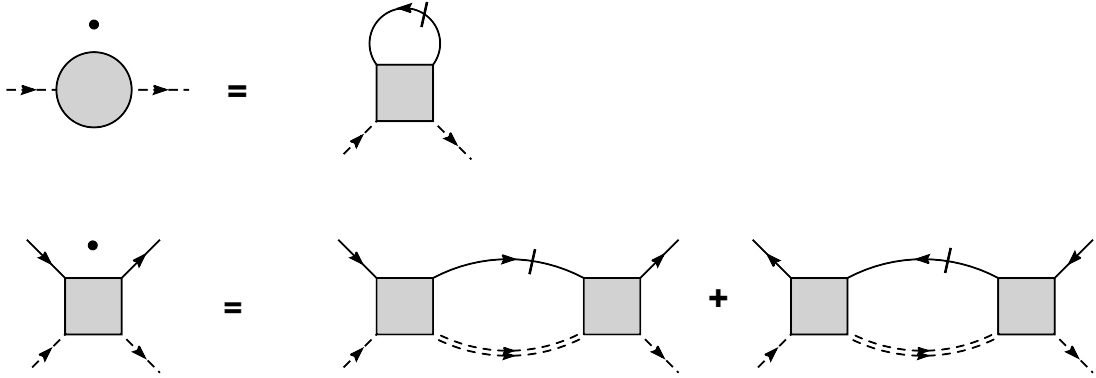}
\caption{Graphical representation of the fRG equation for the self energy $%
\Sigma$ and the two-particle vertex function $\protect\gamma$ as given by
equ. \eqref{frg_equation_sigma} and \eqref{frg_equation_gamma}. The dot on
the l.h.s. represent the derivate w.r.t. $\Lambda$. Solid lines with a
stroke display the lead electrons propagator single scale propagator $%
S^{\Lambda }=G^{0,\Lambda}\frac{d}{d\Lambda}(G^{0,\Lambda})^{-1}G^{0,%
\Lambda} $, whereas the double dashed lines represent the dressed
pseudo-fermion progagator.}
\label{frg_equations_graphic}
\end{figure}

The FRG equations for the self energy and the one-particle irreducible
Green's functions are derived by mapping the generating functional for the
connected Green's functions in Keldysh space (see \cite{Salm,Med,Metz,Jak}
for an introduction into the FRG method) 
\begin{equation*}
W[\eta,\chi]=\log\int D[\psi,\phi]\exp[iS]\exp[(\bar{\psi},\eta)+(\bar{\phi }%
,\chi)+h.c.]
\end{equation*}
with the action 
\begin{equation*}
S=(\bar{\psi},(G^{0})^{-1}\psi)+(\bar{\phi},F_{0}^{-1}\phi)-V[\psi,\phi]
\end{equation*}
where $G_{0}$ and $F_{0}$ are the free propagator Keldysh matrices for the
lead electrons and pseudo fermions, respectively, and $V[\psi,\phi]$ denotes
the exchange interaction, onto a generating functional in which certain high
energy degrees of freedom have been removed at the expense of changing the
interaction part accordingly. We use standard notation for the functional
integrals (see e.g. \cite{Negele}). In particular all time integrals are
extended over the Schwinger double time contour. \newline
Since the interaction is local in space we may integrate out spatial degrees
of freedom. Thus, for stationary situations, the propagators depend only on
a single frequency. We choose to remove the high energy components of the
lead electrons, by introducing a sharp and symmetric---with respect to $%
\mu_{\alpha}$---cutoff $\Lambda$ , expressed by the step function $%
\Theta(|\omega-\mu_{\alpha}|-\Lambda)$, according to%
\begin{equation*}
G^{0,\Lambda}_{\alpha}(\omega)=G^{0}_{\alpha}(\omega)\Theta(|\omega-\mu_{%
\alpha}|-\Lambda)
\end{equation*}
and resulting in a $\Lambda$-dependent generating functional $Z^{\Lambda}$.
It is convenient to perform a Legendre transformation to the generating
functional $W^{\Lambda}$ for the one-particle irreducible (1PI) n-particle
vertex functions $I_{n}^{\Lambda}$ . An expansion of the functional $%
\Gamma^{\Lambda}$ in the fields $I_{n}^{\Lambda}$ leads to the flow
equations for the $I_{n}^{\Lambda}$ , and in particular for the self energy $%
\Sigma$ (we use $\Sigma$ instead of $I_{1}$): 
\begin{equation}
\partial_{\Lambda}\Sigma_{\gamma}^{ba}(\omega)=\frac{1}{2\pi}{\textstyle\sum
\limits_{\alpha,\sigma,c,d}}{\textstyle\sum\limits_{\epsilon=\pm\Lambda}}%
G_{\alpha}^{0,cd}(\epsilon)I_{\gamma a,\gamma b}^{\alpha\sigma
c,\alpha\sigma d}(\epsilon,\omega;\epsilon,\omega)
\label{frg_equation_sigma}
\end{equation}
and for the two-particle vertex function $I\equiv I_{2}:$ %\begin{widetext}
\begin{eqnarray}  \label{frg_equation_gamma}
\partial_{\Lambda} I_{\gamma a, \gamma^{\prime}b}^{\alpha\sigma c,
\alpha^{\prime}\sigma^{\prime}d}(\omega_e,\omega_f;\omega_e^{\prime},%
\omega_f^{\prime}) &=& -\frac{1}{2\pi}\sum_{\epsilon=\pm\Lambda} I_{\gamma a,%
\bar{\gamma}a^{\prime}}^{\alpha\sigma c, \bar{\alpha}\bar{\sigma}c^{\prime}}
G^{0,d^{\prime}c^{\prime}}_{\bar{n}}(\epsilon)
F^{b^{\prime}a^{\prime}}(\omega_e+\omega_f-\epsilon) I_{\bar{\gamma}%
b^{\prime},\gamma^{\prime}b}^{\bar{\alpha}\bar{\sigma} d^{\prime},
\alpha^{\prime}\sigma^{\prime}d}  \notag \\
& &+ I_{\gamma a,\bar{\gamma}a^{\prime}}^{\bar{\alpha}\bar{\sigma}%
d^{\prime}, \alpha^{\prime}\sigma^{\prime}d} G^{0,d^{\prime}c^{\prime}}_{%
\bar{\alpha}}(\epsilon)
F^{b^{\prime}a^{\prime}}(\omega_f-\omega_e^{\prime}+\epsilon) I_{\bar{\gamma}%
b^{\prime},\gamma^{\prime}b}^{\alpha\sigma c, \bar{\alpha}\bar{\sigma}%
c^{\prime}}
\end{eqnarray}
%\end{widetext}
For the sake of simplicity we occasionally omit the frequency arguments of
the vertex function. Keldysh indices are denoted by the latin indices $%
a,b,c,d$. At this stage we have already neglected the term in equ. %
\eqref{frg_equation_gamma} which involves the three particle vertex function 
$I_{3}$ . This term would obey the next RG equation in the infinite
hierarchy of equations for the $I_{n}$ . The principal approximation of our
approach is the neglect of $I_{3}$ and all higher vertex functions. In Fig. %
\ref{frg_equations_graphic} the flow equations \eqref{frg_equation_sigma}
and \eqref{frg_equation_gamma} are depicted graphically. 
%We will later
%argue that these terms introduce corrections to the real part of the self
%energy, which originate from high frequency processes.
%\newline
The boundary conditions at the initial value $\Lambda=\Lambda_{0}=D_{0}$ are
given by 
\begin{align}
\Sigma_{\gamma}^{ba}(\omega) & =0 \\
I_{\alpha a,\alpha^{\prime}b}^{\alpha\sigma
c,\alpha^{\prime}\sigma^{\prime}d}(\omega_{e},\omega_{f};\omega_{e}^{%
\prime},\omega_{f}^{\prime}) & =i\frac{J_{\alpha\alpha^{\prime}}}{4}\vec{\tau%
}_{\sigma\sigma^{\prime}}\vec{\tau}_{\gamma\gamma^{\prime}}\Lambda_{ab}^{cd}%
\,,
\end{align}
independent of frequency. Here, $\Lambda_{ab}^{cd}$ denotes the bare Keldysh
vertex structure, which is given by 
\begin{equation}
\Lambda_{ab}^{cd}=(-1)^{a+1}\delta_{ab}\delta_{cd}\quad\mbox{or}\quad
\Lambda_{ab}^{cd}=\frac{1}{2}(\delta_{ab}\tau_{cd}^{1}+\tau_{ab}^{1}%
\delta_{cd})  \label{bare_vertex_structure}
\end{equation}
in the contour ordered or the Keldysh rotated representation, respectively.
Furthermore we use either a representation of the Keldysh matrices in the
usual Schwinger contour ordered representation ($T,\tilde{T},<,>$) or in
terms of retarded, advanced and Keldysh components \ ($R,A,K$),
respectively. In both representations the matrix Green's functions are given
by \cite{Rammer} 
\begin{equation}
G=%
\begin{pmatrix}
G^{T} & G^{<} \\ 
G^{>} & G^{\tilde{T}}%
\end{pmatrix}
\quad\mbox{or}\quad G=%
\begin{pmatrix}
G^{R} & G^{K} \\ 
0 & G^{A}%
\end{pmatrix}%
\end{equation}
Corresponding representations are used for the pseudo-fermion self energies.
The bare conduction electron local Green's functions at the quantum dot in
both leads are given by \newline
\begin{align*}
G_{\alpha}^{0,R}(\omega) & =-i\pi N_{0}\Theta(D_{0}-|\omega|)=[G_{\alpha
}^{0,A}(\omega)]^{\ast} \\
G_{\alpha}^{0,K}(\omega) & =-2\pi i\tanh\left( \frac{\omega-\mu_{\alpha}}{2T}%
\right) N_{0}\Theta(D_{0}-|\omega|) \\
G_{\alpha}^{0,<}(\omega) & =i2\pi f_{\alpha}(\omega)N_{0}\Theta
(D_{0}-|\omega|) \\
G_{\alpha}^{0,>}(\omega) & =-i2\pi(1-f_{\alpha}(\omega))N_{0}\Theta
(D_{0}-|\omega|)
\end{align*}
where $N_{0}=1/2D_{0}$ is the density of states, $f_{\alpha}(\omega
)=1/(\exp((\omega-\mu_{\alpha})/T)+1)$ is the Fermi function. Since we do
not include a Zeeman term for the conduction electrons, the lead electron
Green functions do not depend on spin and hence the spin indices have been
suppressed. 
%We will take $\Lambda_{0}=D_{0}$ in the following, and use a very
%small value for the imaginary part of the self energy at $\omega=0$ at
%$\Lambda_{0}$.
The dressed pseudo-fermion Green's functions read 
\begin{align*}
F_{\gamma,\lambda}^{R}(\omega) & =\frac{1}{\omega-\Sigma_{\gamma}^{R}(\omega)%
}=[F_{\gamma}^{A}(\omega)]^{\ast} \\
F_{\gamma,\lambda}^{K}(\omega) & =-iA_{\gamma}(\omega)(1-2n_{\gamma,\lambda
}) \\
F_{\gamma,\lambda}^{<}(\omega) & =in_{\gamma,\lambda}A_{\gamma}(\omega) \\
F_{\gamma,\lambda}^{>}(\omega) & =-i(1-n_{\gamma,\lambda})A_{\gamma}(\omega)
\end{align*}
with the spectral function 
\begin{equation}
A_{\gamma}(\omega)=\frac{\Gamma_{\gamma}(\omega)}{[\omega+\alpha\frac{B}{2}%
-R_{\gamma}(\omega)]^{2}+[\Gamma_{\alpha}(\omega)/2]^{2}}\,.
\end{equation}
Here the imaginary part and the real part of the pseudo-fermion self energy
are given by $\Gamma=i(\Sigma^{R}-\Sigma^{A})$ or $R=\mathrm{Re}(\Sigma^{R})$
, respectively. The pseudo-fermion occupation numbers are denoted by $%
n_{\gamma,\lambda}$ . Note that $n_{\gamma,\lambda}\rightarrow0$ in the
limit $\lambda\rightarrow\infty$ (compare the calculation of $\Gamma$ in 
\cite{Paaske1} ). In the presence of a magnetic field, however, it is
necessary to keep the $n_{\gamma,\lambda}$ , as they describe the local spin
polarization $m=\lim_{\lambda\rightarrow\infty}[e^{\lambda/T}(n_{\uparrow
,\lambda}-n_{\downarrow,\lambda})]$.

\section{FRG method applied to the Kondo model}

\label{FRG method applied to the Kondo model}

\subsection{Vertex function structure}

\label{Vertex function structure} Within the perturbative RG method as
formulated in\textit{\ } \cite{pertRG1,pertRG2} only a single component of \ 
$I$ in Keldysh space has been kept. We now consider the structure of the
vertex function $I$ in Keldysh space in more detail. Starting point for the
following considerations is equ. \eqref{frg_equation_gamma} in the Keldysh
rotated form. The leading logarithmic terms in the FRG equation for the
vertex function originate from the product of the Keldysh component of the
lead electron Green function $G^{K}$ with the real part of the retarded or
advanced pseudo fermion propagator $F^{R,A}$. Keeping only these terms in
the sum over Keldysh indices on the right hand side of equ. %
\eqref{frg_equation_gamma} and neglecting the $\Lambda$ dependence in the
energy arguments of the vertex function this equation may be written in a
short hand notation (keeping only Keldysh indices) as

\begin{equation}
\partial _{\Lambda }I_{ab}^{cd}=-\frac{1}{2\pi }\left[ \sum\limits_{a\prime
=1,2}\Pi _{R,A}^{K}I_{aa\prime }^{1d}I_{a\prime b}^{c2}+\sum\limits_{a\prime
=1,2}C_{R,A}^{K}I_{aa\prime }^{c2}I_{a\prime b}^{1d}\right] \,.
\label{RGcoupling}
\end{equation}%
Here $\Pi $ and $C$ denote the kernels in the Peierls and the Cooper
channel, respectively, and the leading components derive from its imaginary
parts. For completeness, however, we state both the imaginary as well as the
real parts. To this end we decompose the individual channels as 
\begin{align}
C_{R,A}^{K}& =\pm C_{r}+iC_{i} \\
\Pi _{R,A}^{K}& =\pm \Pi _{r}+i\Pi _{i}\,,
\end{align}%
with 
\begin{align}
\frac{C_{i}}{2\pi N_{0}}& \equiv -\sum_{\epsilon =\pm \Lambda }\frac{\mathrm{%
tanh}(\frac{\epsilon }{2T})(\omega _{f}+\omega _{e}-\epsilon -\mu _{\bar{%
\alpha}}+\bar{\gamma}\frac{B}{2})}{(\omega _{f}+\omega _{e}-\epsilon -\mu _{%
\bar{\alpha}}+\bar{\gamma}\frac{B}{2})^{2}+\left( \frac{\Gamma _{\gamma }}{2}%
\right) ^{2}}  \label{C_i} \\
\frac{\Pi _{i}}{2\pi N_{0}}& \equiv -\sum_{\epsilon =\pm \Lambda }\frac{%
\mathrm{tanh}(\frac{\epsilon }{2T})({\omega _{f}^{\prime }-\omega
_{e}+\epsilon +\mu _{\bar{\alpha}}+\bar{\gamma}\frac{B}{2})}}{(\omega
_{f}^{\prime }-\omega _{e}+\epsilon +\mu _{\bar{\alpha}}+\bar{\gamma}\frac{B%
}{2})^{2}+\left( \frac{\Gamma _{\gamma }}{2}\right) ^{2}}  \label{Pi_i}
\end{align}%
and 
\begin{align}
\frac{C_{r}}{\pi N_{0}}& \equiv -\sum_{\epsilon =\pm \Lambda }\frac{\mathrm{%
tanh}(\frac{\epsilon }{2T})\,\Gamma _{\gamma }}{(\omega _{f}+\omega
_{e}-\epsilon -\mu _{\bar{\alpha}}+\bar{\gamma}\frac{B}{2})^{2}+\left( \frac{%
\Gamma _{\gamma }}{2}\right) ^{2}} \\
\frac{\Pi _{r}}{\pi N_{0}}& \equiv -\sum_{\epsilon =\pm \Lambda }\frac{%
\mathrm{tanh}(\frac{\epsilon }{2T})\,\Gamma _{\gamma }}{(\omega _{f}^{\prime
}-\omega _{e}+\epsilon +\mu _{\bar{\alpha}}+\bar{\gamma}\frac{B}{2}%
)^{2}+\left( \frac{\Gamma _{\gamma }}{2}\right) ^{2}}\,.
\end{align}%
To simplify matters we have already assumed an energy independent imaginary
part of the pseudo fermion self energy $\Gamma _{\gamma }$ and furthermore
dropped the real part of the self energy. Again we emphasize that $C_{i},\Pi
_{i}\propto \frac{1}{\Lambda }$ and hence produce the leading logarithmic
components, whereas $C_{r},\Pi _{r}\propto \frac{1}{\Lambda ^{2}}$ give rise
to sub-leading corrections. In the following we neglect contributions from $%
C_{r}$ and $\Pi _{r}$ and focus on the leading components only. Hence, we
are led to the equation 
\begin{equation}
\partial _{\Lambda }I_{ab}^{cd}=-\frac{1}{2\pi }\left[ \Pi
_{i}\sum\limits_{a\prime =1,2}I_{aa\prime }^{1d}I_{a\prime
b}^{c2}+C_{i}\sum\limits_{a\prime =1,2}I_{aa\prime }^{c2}I_{a\prime b}^{1d}\,%
\right] .  \label{Keldysh_structure_1}
\end{equation}%
To examine the behaviour under the FRG we perform a Picard-Lindel\"{o}f like
iteration, i.e. we start with the bare vertex structure \ $%
I_{ab}^{cd}=I\Lambda _{ab}^{cd}=\frac{1}{2}I(\delta _{ab}\tau _{cd}^{1}+\tau
_{ab}^{1}\delta _{cd})$ on the right hand side of equ. %
\eqref{Keldysh_structure_1}. By performing the sum over Keldysh indices we
see that in both the Cooper and Peierls channels this structure is
reproduced. Hence, in leading order there is only one invariant amplitude,
of the structure of the bare vertex 
\begin{equation}
I_{ab}^{cd}\sim I\Lambda _{ab}^{cd}\,.  \label{single_component}
\end{equation}%
confirming the treatment in\textit{.} \cite{pertRG1,pertRG2}. To recover the
perturbative RG scheme of \cite{pertRG1, pertRG2} from the FRG equation %
\eqref{frg_equation_gamma} the following additional steps have to be
performed: As a first step the imaginary part of the pseudo fermion self
energy is neglected in the denominator of equ. \eqref{C_i} and \eqref{Pi_i}.
Moreover the energies of the pseudo fermions are put on-shell, i.e. $\omega
_{f}=-\gamma B/2$. Then the coupling function depends only on a single
frequency, the frequency of the incoming electron.. Finally the cutoff
dependences in the Cooper and Peierls channels (equ. \eqref{C_i} and %
\eqref{Pi_i}) are approximated by a window function 
\begin{equation}
\sum_{\epsilon =\pm \Lambda }\frac{\mathrm{sign}(\epsilon )}{\epsilon
-\Delta \omega }=2\frac{\Lambda }{\Lambda ^{2}-(\Delta \omega )^{2}}\approx
2\Theta (\Lambda -|\Delta \omega |)  \label{pertRG_theta}
\end{equation}%
as described in more detail in \cite{pertRG1,pertRG2}. Here $\Delta \omega $
depends on $V,B$ and the incoming and outgoing frequencies. Finally, in
order to incorporate decoherence effects, which serve as a cutoff for the RG
flow, a relaxation rate $\Gamma $ , which has contributions from both, self
energy and vertex corrections, is incorporated in equ. \eqref{pertRG_theta}
by replacing $\Theta (\Lambda -|\Delta \omega |)$ by $\Theta (\Lambda -\sqrt{%
(\Delta \omega )^{2}+\Gamma ^{2}})$, where $\Gamma $ in turn is calculated
by a second order golden rule expression with renormalized vertices. These
two equations (i.e. for the coupling function and the decoherence rate $%
\Gamma $) are then solved iteratively, until convergence is reached.

In the following we will show how the decoherence rate (to be more precise
the imaginary part of the pseudo fermion self energy, which is part of the
decoherence rate) arises in a natural way. To this end we first simplify the
coupled equations \eqref{frg_equation_sigma} and \eqref{frg_equation_gamma}
in an appropriate way and then solve these equations for different parameter
regimes.

\subsection{Approximated flow equations}

In order to simplify the \textit{coupled} differential equations for the
imaginary part of the pseudo fermion self energy $\Gamma $ and the effective
interaction $I$ we first neglect the frequency dependence of the leading
component of \ $I$ ,%
\begin{equation}
\partial _{\Lambda }I_{\gamma ;\gamma ^{\prime }}^{\alpha \sigma ;\alpha
^{\prime }\sigma ^{\prime }}=-\frac{1}{2\pi }\sum_{\bar{\alpha}\bar{\gamma}%
\bar{\sigma}}\Big[~I_{\gamma ;\bar{\gamma}}^{\alpha \sigma ,\bar{\alpha}\bar{%
\sigma}}I_{\bar{\gamma},\gamma ^{\prime }}^{\bar{\alpha}\bar{\sigma};\alpha
^{\prime }\sigma ^{\prime }}~C_{\bar{\gamma}}^{\bar{\alpha}}(\Lambda
)+~I_{\gamma ;\bar{\gamma}}^{\bar{\alpha}\bar{\sigma},\alpha \sigma }I_{\bar{%
\gamma},\gamma ^{\prime }}^{\alpha \sigma ;\bar{\alpha}\bar{\sigma}}\Pi _{%
\bar{\gamma}}^{\bar{\alpha}}(\Lambda )\Big]\,,
\label{approx_leading_component}
\end{equation}%
where 
\begin{align}
C_{\bar{\gamma}}^{\bar{\alpha}}(\Lambda )& =-2\pi N_{0}\sum_{\epsilon =\pm
\Lambda }\frac{\mathrm{tanh}(\frac{\epsilon }{2T})(-\epsilon -\mu _{\bar{%
\alpha}}+\bar{\gamma}\frac{B}{2})}{(-\epsilon -\mu _{\bar{\alpha}}+\bar{%
\gamma}\frac{B}{2})^{2}+\left( \frac{\Gamma _{\gamma }}{2}\right) ^{2}}
\label{approx_cooper} \\
\Pi _{\bar{\gamma}}^{\bar{\alpha}}(\Lambda )& =-2\pi N_{0}\sum_{\epsilon
=\pm \Lambda }\frac{\mathrm{tanh}(\frac{\epsilon }{2T})({\epsilon +\mu _{%
\bar{\alpha}}+\bar{\gamma}\frac{B}{2})}}{(\epsilon +\mu _{\bar{\alpha}}+\bar{%
\gamma}\frac{B}{2})^{2}+\left( \frac{\Gamma _{\gamma }}{2}\right) ^{2}}\,.
\label{approx_peierls}
\end{align}%
The RG equation for the imaginary part of the pseudo fermion self energy $%
\Gamma =i(\Sigma ^{R}-\Sigma ^{A})$ requires knowledge of the vertex
function $I_{\gamma a,\gamma b}^{\alpha ,\sigma c;\alpha \sigma d}(\epsilon
,0;\epsilon ,0)$ at $\epsilon =\pm \Lambda $. By simply replacing this
vertex function by an energy independent one of the form %
\eqref{single_component} the imaginary part of the pseudo fermion self
energy is seen to vanish identically. Here it is necessary to take into
account the subleading Keldysh components of $I$ . To this end we use the
leading components on the right hand side of the equation for $I(\epsilon
,0;\epsilon ,0)$ , equ. \eqref{RGcoupling}, to generate the subleading
Keldysh components, which we denote by $\widetilde{I}(\epsilon )$ : 
\begin{equation}  \label{approx_gamma_tilde}
\partial _{\Lambda }\widetilde{I}_{\gamma a;\gamma b}^{\alpha \sigma
c;\alpha \sigma d}(\epsilon )=-\frac{1}{2\pi }\Big[I_{\gamma ;\bar{\gamma}%
}^{\alpha \sigma ,\bar{\alpha}\bar{\sigma}}~I_{\bar{\gamma};\gamma }^{\bar{%
\alpha}\bar{\sigma};\alpha \sigma }~\Lambda _{aa^{\prime }}^{cc^{\prime
}}\Lambda _{b^{\prime }b}^{d^{\prime }d}\tilde{C}_{a^{\prime }b^{\prime }%
\bar{\gamma}}^{c^{\prime }d^{\prime }\bar{\alpha}}(\Lambda ,\epsilon
)+I_{\gamma ;\bar{\gamma}}^{\bar{\alpha}\bar{\sigma},\alpha \sigma }~I_{\bar{%
\gamma};\gamma }^{\alpha \sigma ;\bar{\alpha}\bar{\sigma}}~\Lambda
_{aa^{\prime }}^{d^{\prime }d}\Lambda _{b^{\prime }b}^{cc^{\prime }}\tilde{%
\Pi}_{a^{\prime }b^{\prime }\bar{\gamma}}^{c^{\prime }d^{\prime }\bar{\alpha}%
}(\Lambda ,\epsilon )\Big]
\end{equation}%
where the Einstein summation convention is used. The kernels in the Cooper
and Peierls channel read 
\begin{align}
\tilde{C}_{a^{\prime }b^{\prime }\bar{\gamma}}^{c^{\prime }d^{\prime }\bar{%
\alpha}}(\Lambda ,\epsilon )& =\sum_{\tilde{\epsilon}=\pm \Lambda }G_{\bar{%
\alpha}}^{d^{\prime }c^{\prime }}(\tilde{\epsilon})F_{\bar{\gamma}%
}^{b^{\prime }a^{\prime }}(\epsilon -\tilde{\epsilon}) \\
\tilde{\Pi}_{a^{\prime }b^{\prime }\bar{\gamma}}^{c^{\prime }d^{\prime }\bar{%
\alpha}}(\Lambda ,\epsilon )& =\sum_{\tilde{\epsilon}=\pm \Lambda }G_{\bar{%
\alpha}}^{d^{\prime }c^{\prime }}(\tilde{\epsilon})F_{\bar{\gamma}%
}^{b^{\prime }a^{\prime }}(\epsilon +\tilde{\epsilon})\,.
\end{align}%
With the aid of \eqref{approx_gamma_tilde} the equation for the imaginary
part of the pseudo fermion self energy may be cast in the form 
\begin{equation}
\partial _{\Lambda }\Gamma _{\gamma }=-\frac{i}{2\pi }\sum_{\epsilon =\pm
\Lambda }\left( \widetilde{I}_{\gamma 1;\gamma 1}^{\alpha \sigma c;\alpha
\sigma d}(\epsilon )-\widetilde{I}_{\gamma 2;\gamma 2}^{\alpha \sigma
c;\alpha \sigma d}(\epsilon )\right) G_{\alpha }^{cd}(\epsilon )
\label{approx_Gamma}
\end{equation}%
Hence equ. \eqref{approx_leading_component}, \eqref{approx_gamma_tilde} and %
\eqref{approx_Gamma} provide the desired system of coupled differential
equations for the vertex function and the imaginary part of the pseudo
fermion self energy. %Note that all these equations are given in the
%Keldysh rotated representation. Optionally we may switch to use
%contour ordered representation for equ. \eqref{approx_gamma_tilde} and \eqref{approx_Gamma}.
%In this case equ. \eqref{approx_gamma_tilde} keeps it structure but with
%the replacement of the bare vertex structure from the Keldysh rotated
%form to the contour ordered one (cf. equ. \eqref{bare_vertex_structure} ). Additionally
%the propagators have to be changed. The equation for the imaginary part of the
%self energy changes according to
%\be\label{approx_Gamma_2}
%\dot{\Gamma}_{\gamma} = i\dot{\Sigma}^>_{\gamma} = -\frac{i}{2\pi}
%\sum_{\epsilon=\pm\Lambda}
%\tilde{\gamma}_{\gamma 1;\gamma 2}^{\alpha\sigma c;\alpha\sigma d}(\epsilon)
%G_{\alpha}^{cd}(\epsilon)
%\ee
%Subsequently we consider different parameter regimes, in particular for
%the temperature $T$ and the bias voltage $V$.\\
For the subsequent analysis we are going to use dimensionless couplings $%
g=N_{0}I$. Additionally, the spin dependence of the vertex function is
parameterized as $g_{\gamma \gamma ^{\prime }}^{\alpha \sigma ,\alpha
^{\prime }\sigma ^{\prime }}=g_{\alpha \alpha ^{\prime }}\mathbf{\tau }%
_{\sigma \sigma ^{\prime }}\cdot \mathbf{\tau }_{\gamma \gamma ^{\prime }}$
and, for the sake of simplicity we restrict ourselves to the case of
symmetric couplings $g_{\alpha \alpha ^{\prime }}=g$. Furthermore, in order
to establish a more direct connection to Poor man's RG schemes \cite%
{pertRG1,pertRG2,And} and to prevent a factor of $i$ from pervading the
formulas, we employ a redefinition of $g$, i.e. $g\rightarrow i8g$, which
results in the new boundary condition $g=g_{0}=N_{0}J$ at $\Lambda
_{0}=D_{0} $.

\subsection{Weak coupling and linear response at finite temperature}

\label{Weak coupling and linear response at finite temperature} 
\begin{figure}[tbp]
\begin{minipage}{65mm}
\includegraphics[width=0.97 \linewidth]{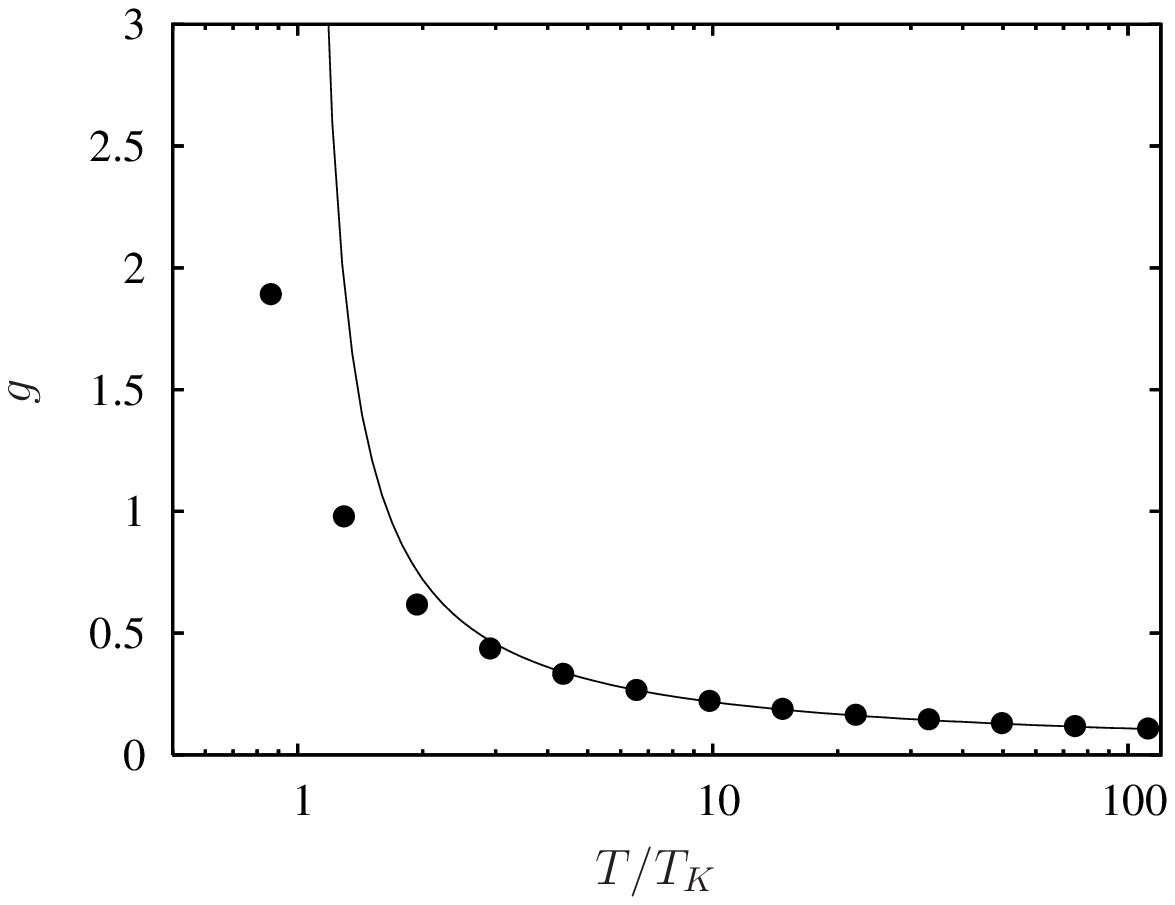}
\caption{Leading component of the coupling function $g(0)$ as a
function of $T/T_{K}$ for values down to $T/T_{K}\approx1$. Dots: results from
FRG; solid line: $g=\frac{1}{2\log(T/T_{K})}$.}
\label{weak_coupling_g_T}
\end{minipage}
\hfil
\begin{minipage}{65mm}
\includegraphics[width=\linewidth]{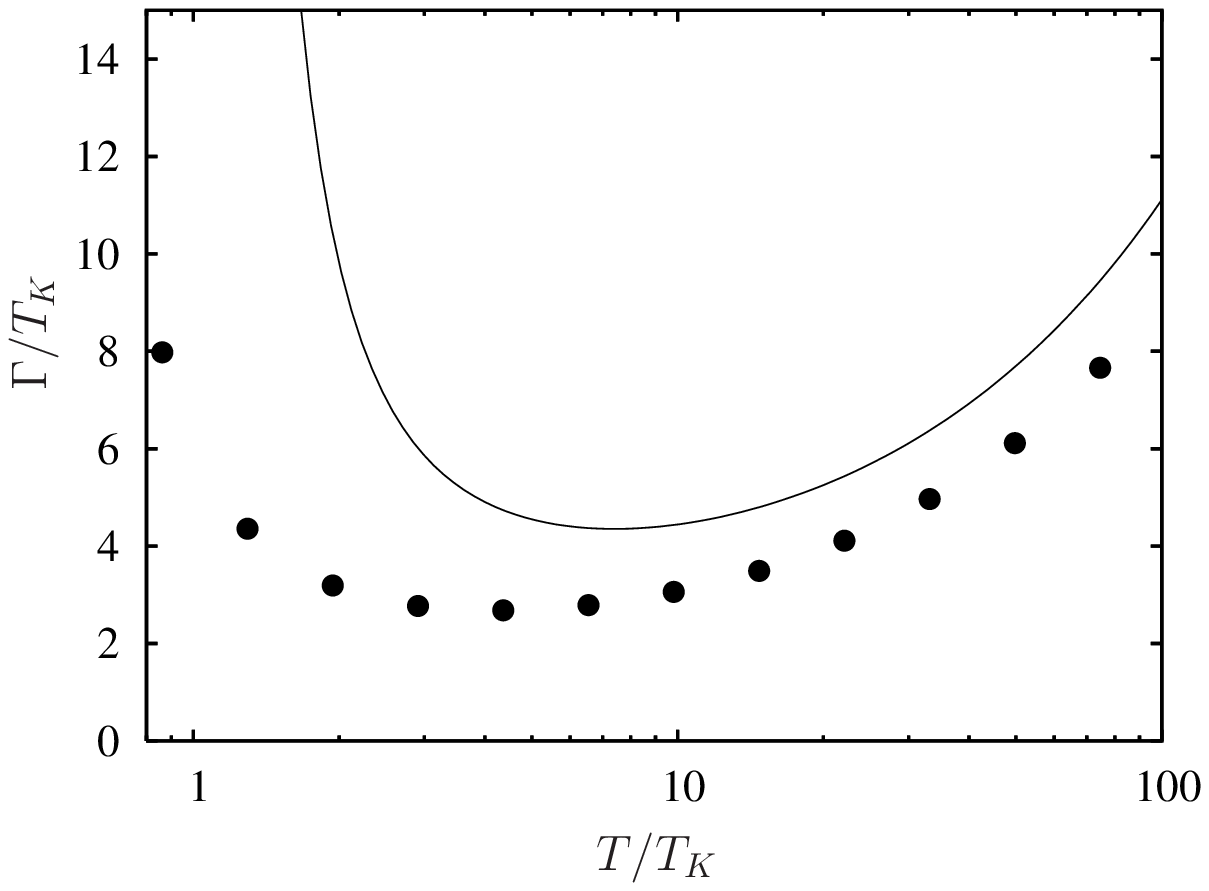}
\caption{ Imaginary part of the self energy
$\Gamma(0)$ versus $T/T_{K}$ for values down to $T/T_{K}\approx1$. Dots:
numerical results from FRG; solid line: $\Gamma=3\pi T\frac{1}{(2\log
(T/T_{K}))^{2}}$.}
\label{weak_coupling_Gamma_T}
\end{minipage}
\end{figure}

To begin
with a quantitative analysis we focus on the weak coupling regime $%
B,V\rightarrow 0$ and $T/T_{K}\gg 1$, where the Kondo temperature is defined
by $T_{K}=D_{0}\exp (-1/(2g_{0}))$. In this case the Cooper and Peierls
kernels \eqref{approx_cooper} and \eqref{approx_peierls} simplify to 
\begin{equation}
C(\Lambda )=4\pi N_{0}\frac{\Lambda \,\mathrm{tanh}(\frac{\Lambda }{2T})}{%
\Lambda ^{2}+\left( \Gamma _{\gamma }/2\right) ^{2}}=-\Pi (\Lambda )\,.
\end{equation}%
Performing the internal spin sums in equ. \eqref{approx_leading_component} ,
one finds 
\begin{equation}
\partial _{\Lambda }g=-2g^{2}\tanh \left( \frac{\Lambda }{2T}\right) \frac{%
\Lambda }{\Lambda ^{2}+\Gamma ^{2}}\,,  \label{weak_coupling_T_g}
\end{equation}%
Similarly, in the absence of a magnetic field the RG equations for $\Gamma $
and $\widetilde{g}$ take the simpler form 
%Likewise for the imaginary part of the pseudo fermion self energy and
%the vertex part $\tilde{\gamma}(\epsilon)$%
\begin{align} \label{weak_coupling_T_g_tilde} 
\partial _{\Lambda }\widetilde{g}(\epsilon )& =3g^{2}\sum_{\tilde{\epsilon}%
=\pm \Lambda }f(-\tilde{\epsilon})\frac{\Gamma }{(\epsilon -\tilde{\epsilon}%
)^{2}+\left( \Gamma /2\right) ^{2}} \\\label{weak_coupling_T_Gamma}
\partial _{\Lambda }\Gamma & =\sum_{\epsilon =\pm \Lambda }f(\epsilon )%
\tilde{g}(\epsilon )\,.
\end{align}%
In this case the corresponding initial conditions read 
\begin{equation}
\widetilde{g}^{\Lambda _{0}=D}(\epsilon )=0\qquad \mbox{and}\quad \Gamma
^{\Lambda _{0}=D}=0\,.
\end{equation}%
For the numerical treatment of these equations, however, we have to take a
small but finite value for $\Gamma ^{\Lambda _{0}}$ . It turns out that the
solution of the FRG equations does not depend much on the choice of the
initial value of $\Gamma ^{\Lambda _{0}}$ as long as it is taken to be
sufficiently small. \newline
We now solve the FRG equations for values of $T/T_{K}\gg 1$ and obtain
results shown in Fig. \ref{weak_coupling_g_T} and Fig. \ref{weak_coupling_Gamma_T}. 
In Fig. \ref{weak_coupling_g_T} the leading Keldysh
component of the coupling at the Fermi energy, $g(\omega =0)$, is shown at
zero bias as a function of $T/T_{K}.$ At temperatures $T\gg T_{K}$ the
numerical result (dots) agrees very well with the result of renormalized
perturbation theory, $g=1/2\ln (T/T_{K})$ . It is seen that the effect of
spin relaxation leads to a reduction of the growth of $g(\omega =0)$ as
compared to the weak coupling result, as expected, but does not cause $%
g(\omega =0)$ to take a temperature independent value of order unity at $%
T\ll T_{K}$ . The relaxation rate $\Gamma $ , shown in Fig. \ref%
{weak_coupling_Gamma_T}, also agrees well with the renormalized perturbation
theory expression, $\Gamma =3\pi T/[2\ln (T/T_{K})]^{2}$ at $T\gg T_{K}$. At
lower temperatures $\Gamma $ is seen to fall well below the diverging
perturbative result, as expected, but does not appear to approach a limiting
constant value of order $T_{K}$ at $T\ll T_{K}$ .

\begin{figure}[tbp]
\begin{minipage}{65mm}
\vspace{5mm}
\includegraphics[width=\linewidth]{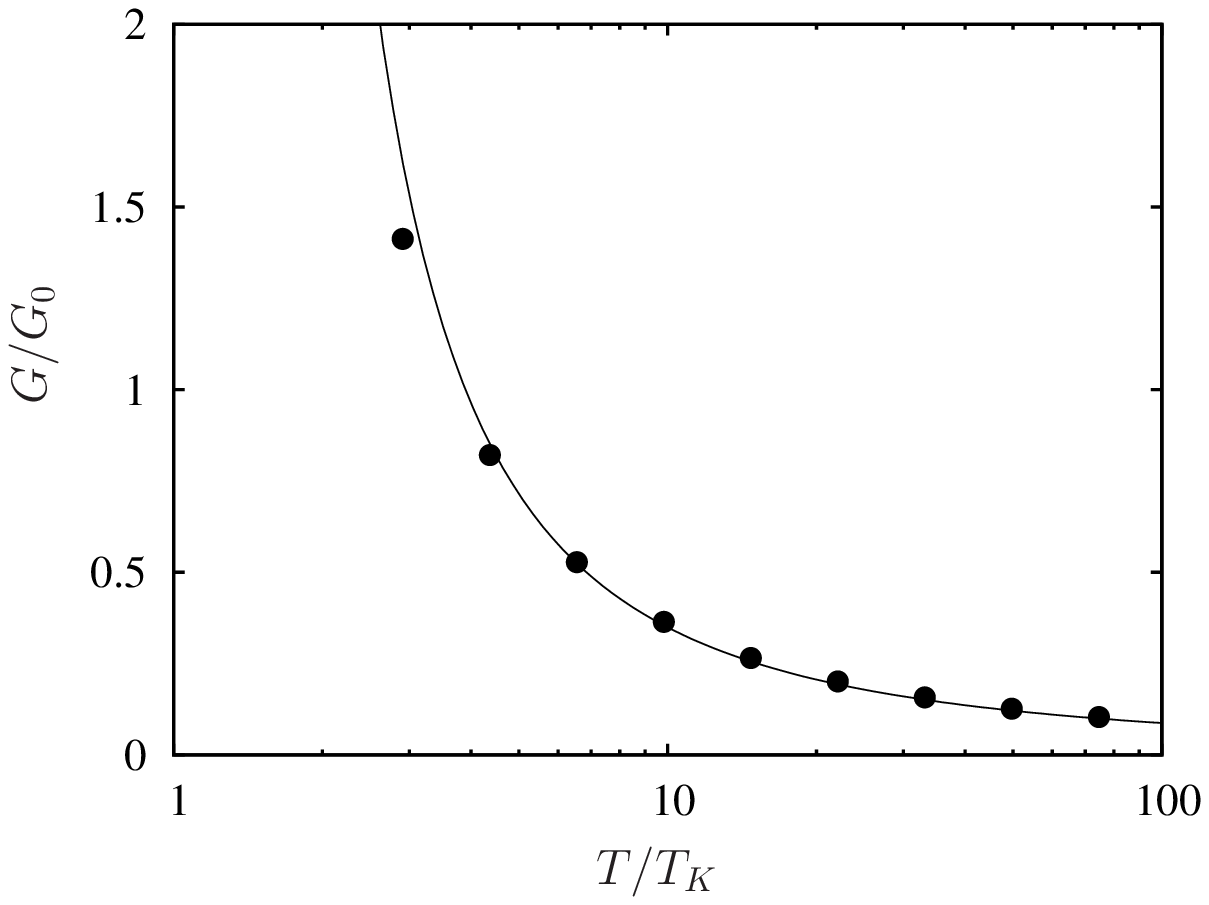}\caption[Equilibrium:
conductance]{ Conductance in units of $G_{0}=2e^{2}/h$ versus $T/T_{K}$. Dots:
numerical results from FRG; solid line: $G/G_{0}=\frac{3\pi^{2}}{16}\frac
{1}{\log^{2}(T/T_{K})}$. }%
\label{weak_coupling_G_T}%
\end{minipage}
\hfil
\begin{minipage}{65mm}
\includegraphics[width=\linewidth]{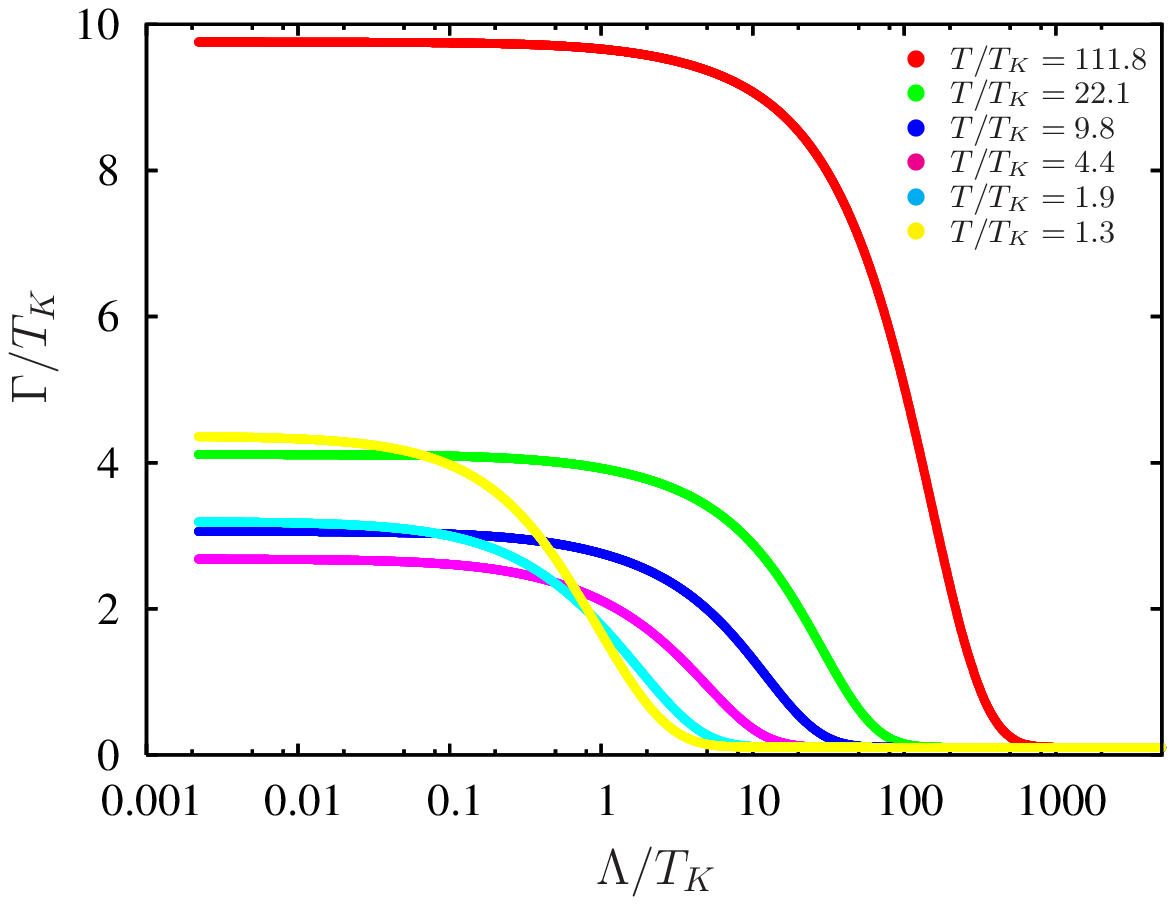}\caption[Equilibrium:
imaginary part of the self energy ]{ FRG-flow of $\Gamma(\Lambda)$ for various
values of $T/T_{K}$.}%
\label{weak_coupling_Gamma_T_flow}%
\end{minipage}
\end{figure}

The conductance $G$ (in units of $e^{2}/\pi \hbar $ ), in the absence of a
magnetic field, may be approximately obtained from the golden rule
expression 
\begin{equation}
G(T)=\frac{3\pi }{4}g^{2}(\omega =0)  \label{conductance_expression}
\end{equation}%
Fig. \ref{weak_coupling_G_T} shows a comparison of the linear conductance as
obtained from equ. \eqref{conductance_expression} (dots) and the weak
coupling result (solid line) $G=(3\pi /16)/[\ln (T/T_{K})]^{2}$ . Even
though the rapid rise of $G$ as $T$ approaches $T_{K}$ is somewhat slowed
down by the effect of the relaxation rate $\Gamma $ , the slowdown is too
weak or happens too late so that $G$ rises beyond the unitarity limit. 
%%%%%%%%%%%%%%%%%%%%%%%%%%%%%%%%%%%%%%%%%%%%%%%%%%%%%%%%%%%%%%%%%%%%%
%%%%%%%%%%%%%%%%%%%%%%%%%%%%%%%%%%%%%%%%%%%%%%%%%%%%%%%%%%%%%%%%%%%%%%
%%%%%%%%%%%%%%%%%%%%%%%%%%%%%%%%%%%%%%%%%%%%%%%%%%%%%%%%%%%%%%%%%%%%%

It is somewhat surprising that the initial strong growth of $g$ is not cut
off more effectively by $\Gamma $ . After all, at $T\gtrsim T_{K}$ the
relaxation rate reaches values of $\Gamma \simeq 3T_{K}$ so that it might be
expected to be more important than the temperature in cutting off the RG
flow. The reason for this different behavior is apparent from Fig. \ref%
{weak_coupling_Gamma_T_flow} which shows the behavior of $\Gamma $ as a
function of the RG cutoff scale $\Lambda $ : A sizeable value of $\Gamma $
is only generated at the scale $\Lambda \lesssim T$ and not at $T_{K}$ as
one might have expected. Therefore, at $T\ll T_{K}$ and $\Lambda \simeq
T_{K} $ the relaxation width $\Gamma $ is still exponentially small and does
not serve to cut off the flow of $g$ to large values, which explains why $%
g,\Gamma ,$and $G$ (at $\Lambda =0)$ appear to grow beyond the unitarity
limit. We emphasize again that the above results depend sensitively on the
approximation of neglecting the energy dependence, both, of the coupling
function $g(\omega )$ and the relaxation rate $\Gamma (\omega )$ (see below).

\subsection{Weak coupling at finite bias voltage}

\begin{figure}[tbp]
\begin{minipage}{65mm}
\includegraphics[width=\linewidth]{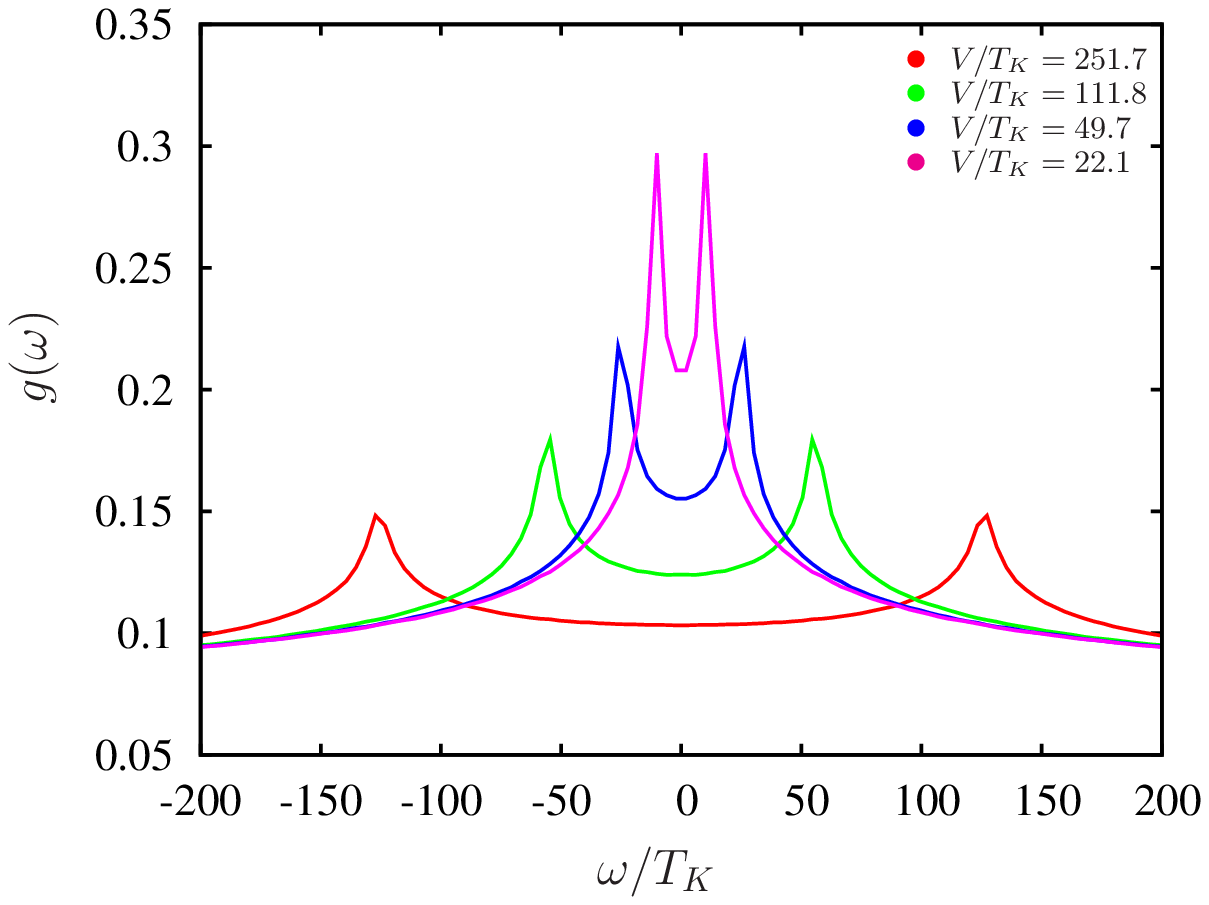} \caption{ Leading
component of the coupling function $g(\omega)$ versus $\omega/T_{K}$ for various
values of $V/T_K$.}%
\label{weak_coupling_g_V}
\end{minipage}
\hfil
\begin{minipage}{65mm}
\vspace{5mm}
\includegraphics[width=\linewidth]{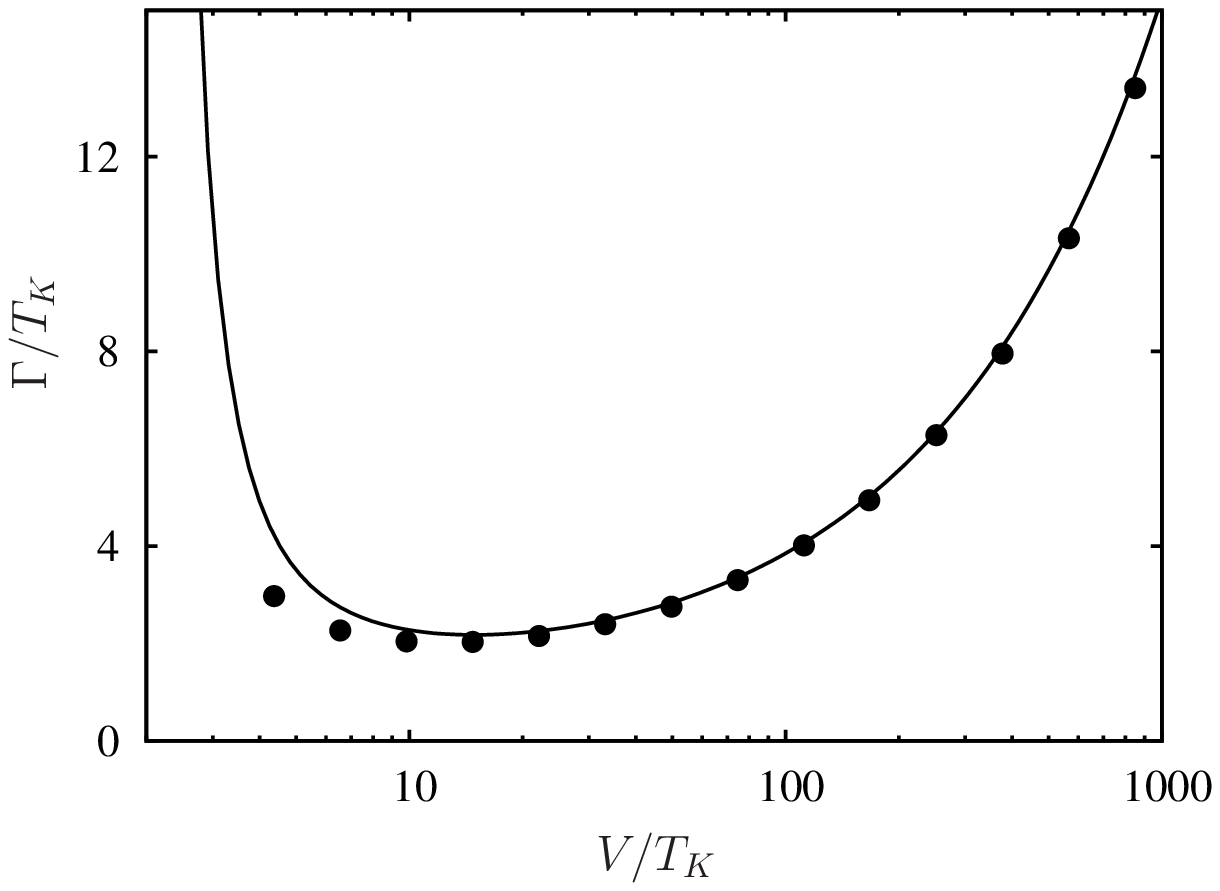} \caption{Imaginary
part of the self energy $\Gamma$ versus $V/T_{K}$. Dots: numerical results
from FRG; solid line: $\Gamma=\frac{3\pi}{4} V\frac{1}{(2\log(V/2T_{K}))^{2}}%
$. }%
\label{weak_coupling_Gamma_V}%
\end{minipage}
\end{figure}

As it is known \cite{pertRG1,pertRG2} that in case of a finite bias voltage $%
V$ the frequency dependence of the coupling function is crucial, in the
following analysis we follow \cite{pertRG1,pertRG2} and maintain the
frequency of the lead electron. We approximate the kernels in the Peierls
and Cooper channels by using 
\begin{equation}
\sum_{\epsilon =\pm \Lambda }\tanh (\frac{\epsilon }{2T})\frac{\epsilon
-\Delta \omega }{(\epsilon -\Delta \omega )^{2}+(\Gamma /2)^{2}}\approx
2\tanh (\frac{\Lambda }{2T})\Theta (\Lambda -\sqrt{(\Delta \omega
)^{2}+(\Gamma /2)^{2}})\,
\end{equation}%
where $\Delta \omega =\omega -\mu _{\alpha }$. Hence, the equations for the
leading component of the vertex function, $g(\omega )$ , the subleading
vertex component $\tilde{g}$ , and the imaginary part of the pseudo fermion
self energy $\Gamma $ take the form 
\begin{align}
\partial _{\Lambda }g(\omega )& =-g^{2}(\omega )\frac{1}{\Lambda }%
\sum_{\alpha }\Theta (\Lambda -\sqrt{(\omega -\mu _{\alpha })^{2}+(\Gamma
/2)^{2}})  \label{weak_coupling_V_g} \\
\partial _{\Lambda }\tilde{g}(\epsilon )& =\frac{3}{2}g^{2}(0)\sum_{\tilde{%
\epsilon}=\pm \Lambda }\sum_{\alpha }f_{\alpha }(-\tilde{\epsilon})\frac{%
\Gamma }{(\epsilon -\tilde{\epsilon})^{2}+\left( \Gamma /2\right) ^{2}} \\
\partial _{\Lambda }\Gamma & =\frac{1}{2}\sum_{\epsilon =\pm \Lambda
}\sum_{\alpha }f_{\alpha }(\epsilon )\tilde{g}(\epsilon )\,,
\end{align}%
where 
\begin{equation*}
f_{\alpha }(\epsilon )=\frac{1}{\exp \left( \frac{\epsilon -\mu _{\alpha }}{T%
}\right) +1},\qquad \quad \mu _{\alpha }=\pm eV/2\,.
\end{equation*}%
Results of a numerical solution of these RG equations are shown in Fig. \ref%
{weak_coupling_g_V} and Fig. \ref{weak_coupling_Gamma_V}. In Fig. \ref%
{weak_coupling_g_V} we display $g(\omega )$ for various values of $V/T_{K}$.
Fig. \ref{weak_coupling_Gamma_V} shows the imaginary part of the pseudo
fermion self energy $\Gamma $ (dots) with the approximate expression $\Gamma
=\frac{3\pi }{4}V\frac{1}{\log ^{2}(V/2T_{K})}$ (solid line), which is seen
to agree very well in the regime $V\gg T_{K}$ .\newline
At finite voltage the charge current $I_{c}$ is approximately given by 
%\begin{widetext}

\begin{equation}
I_{c}=\frac{3\pi ^{2}e}{2\hbar }\int \frac{d\omega }{2\pi }\int \frac{%
d\omega \prime }{2\pi }\int \frac{d\epsilon }{2\pi }|g(\omega ,\epsilon
;\omega \prime ,\epsilon +\omega -\omega \prime )|^{2}[f_{L}(\omega
)-f_{R}(\omega )]A(\epsilon )A(\epsilon +\omega -\omega \prime )\,\,..
\label{current}
\end{equation}%
$.$\bigskip

%\end{widetext}
In the limit of vanishing relaxation rate $\Gamma $ , the above expression
simplifies to%
\begin{equation}
I_{c}=\frac{3\pi e}{4\hbar }\int d\omega |g(\omega ,0;\omega
,0)|^{2}[f_{L}(\omega )-f_{R}(\omega )]\,.  \label{current_wc}
\end{equation}%
\bigskip

\begin{figure}[h]
\centering
\includegraphics[width=0.45\textwidth]{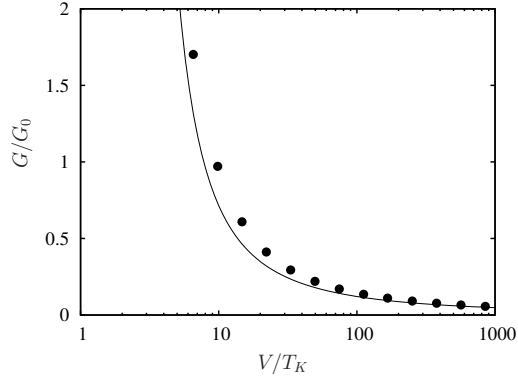}
\caption{Conductance $G/G_{0}=I/V$ versus $V/T_{K}$. Dots: numerical result
from FRG; solid line: $G/G_{0}=\frac{3\protect\pi ^{2}}{16}\frac{1}{\log
^{2}(V/2T_{K})}$. }
\label{weak_coupling_G_V}
\end{figure}

\begin{figure}[tbp]
\begin{minipage}{65mm}
\vspace{0mm}
\includegraphics[width=\linewidth]{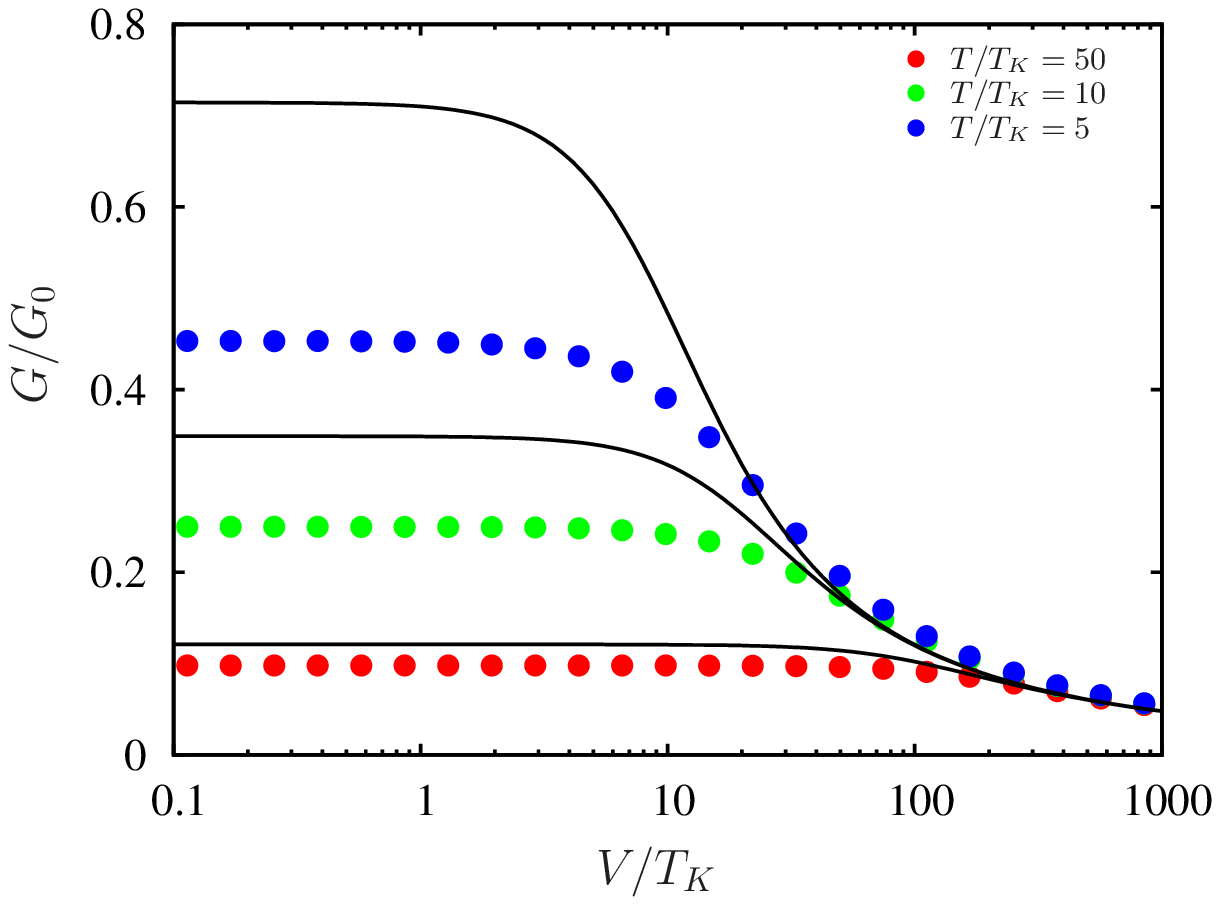}\caption[Nonequilibrium:
finite temperature]{ Conductance in units of $G_{0}=2e^{2}/h$ as a function of
$V/T_{K}$. Dots denote the result obtained by solving the fRG equations
numerically. The solid line corresponds to the weak coupling form
$G/G_{0}=3\pi^{2}1/16\log^{2}\left(  \frac{\sqrt{(V/2)^{2}+T^{2}}}{T_{K}%
}\right)  $. }%
\label{weak_coupling_G_V_T}%
\end{minipage}
\hfil
\begin{minipage}{65mm}
\vspace{-12mm}
\includegraphics[width=\linewidth]{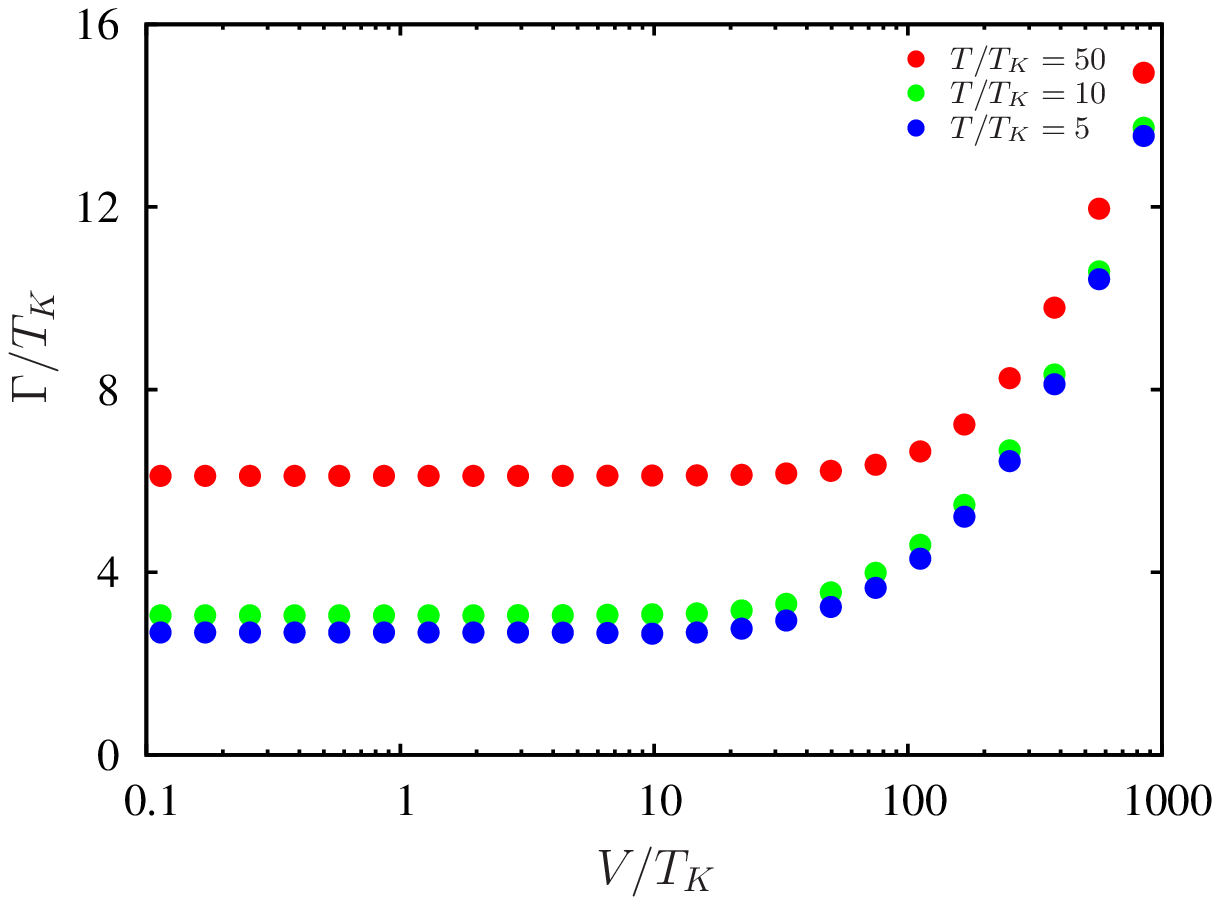}\caption[Nonequilibrium:
finite temperature]{ Imaginary part of the self energy $\Gamma$ as a function
of $V/T_{K}$. Dots represent again the results from the fRG.}%
\label{weak_coupling_Gamma_V_T}%
\end{minipage}
\end{figure}

In Fig. \ref{weak_coupling_G_V} we show the conductance $G=I_{c}/V$ \ in
units of $G_{0}$ as obtained from equ. \eqref{current_wc} versus $V/T_{K}$
at $T=0$. Again, dots represent the numerical evaluation of the FRG
equations, whereas the solid line represents an approximate analytical form $%
G/G_{0}=\frac{3\pi ^{2}}{16}\frac{1}{\ln ^{2}(V/2T_{K})}$. One obtains this
analytical result by replacing $g(\omega )$ by $g(\omega =0)$ in equ. %
\eqref{weak_coupling_V_g}. However, because of the distinctive form of $%
g(\omega ),$ which exhibits pronounced peaks at the resonance frequencies $%
\omega =\pm V/2$ the resulting $G(V)/G_{0}$ is slightly enhanced compared to
the analytical form. %
%
%
%%%%%%%%%%%%%%%%%%%%%%%%%%%%%%%%%%%%%%%%%%%%%%%%%%%%%%%%%%%%%%%%%%%%%%%%
%%%%%%%%%%%%%%%%%%%%%%%%%%%%%%%%%%%%%%%%%%%%%%%%%%%%%%%%%%%%%%%%%%%%%%%%
%%%%%%%%%%%%%%%%%%%%%%%%%%%%%%%%%%%%%%%%%%%%%%%%%%%%%%%%%%%%%%%%%%%%%%%%
%
%
%
In Fig. \ref{weak_coupling_G_V_T} results on the conductance $G/G_{0}$
versus $V/T_{K}$ are presented for temperatures $T/T_{K}=50,10,5$ , such
that one remains in the perturbative regime even in the limit $V\rightarrow
0 $. Dots are the FRG results, whereas the solid line represents $G/G_{0}=%
\frac{3\pi ^{2}}{4}g^{2}$ with the approximate vertex function 
\begin{equation}
g(\omega =0)\approx \frac{1}{2\ln \frac{\sqrt{(V/2)^{2}+T^{2}}}{T_{K}}}\,.
\end{equation}%
%
%
%
%
%
%%%%%%%%%%%%%%%%%%%%%%%%%%%%%%%%%%%%%%%%%%%%%%%%%%%%%%%%%%%%%%%%%%%%%%%%%%%%%%%%%%%%%%%%%%%
%%%%%%%%%%%%%%%%%%%%%%%%%%%%%%%%%%%%%%%%%%%%%%%%%%%%%%%%%%%%%%%%%%%%%%%%%%%%%%%%%%%%%%%%%%%
%
%
%
The frequency dependence of $g(\omega )$ together with the integration over
the difference of the Fermi functions in the left and right lead in equ. %
\eqref{current_wc} lead to a distinct discrepancy between the FRG result and
the analytical form for $V\lesssim T$. In Fig. \ref{weak_coupling_Gamma_V_T}
results on $\Gamma $ versus $V/T_{K}$ in the same temperature regime as in
Fig. \ref{weak_coupling_G_V_T} are shown. As can be seen, in the temperature
range $3K\lesssim T\lesssim 10K$ $\Gamma $ saturates to a value of the order
of $T_{K}$ when the voltage falls below the corresponding temperature.

\subsection{Towards the strong coupling regime}

From the previous considerations, i.e. the analysis and the solution of the
approximated set of FRG equations \ref{weak_coupling_T_g}, \ref{weak_coupling_T_g_tilde} and \ref{weak_coupling_T_Gamma} it can be inferred
that this set of equations in its present form is not appropriate to capture
the physics in the strong coupling regime, i.e. in the regime $T,V\ll T_{K}$%
. There are two major deficiencies in the approximate RG equations: (1) the
unitarity limit on the conductance, $G\leqslant G_{0}$ is not respected; (2)
the spin relaxation rate $\Gamma $ is found to rise with decreasing
temperature/voltage in the strong coupling regime, whereas $\Gamma $ is
expected to saturate to a value of order $T_{K}$ . There might be several
reasons responsible for this failure.

First, the effect of higher order correlation functions is neglected in the
approximation scheme discussed so far. The additional terms in the RG
equation for the coupling function $g(\omega =0)$ generated thereby are
similar, but not identical, to the higher order loop corrections to the poor
man's scaling approach, and may be expected to slow down the growth of the
coupling near the energy scale $T_{K}$ . A quantitative control of these
terms within FRG appears to be out of reach. Nonetheless, in the following
we shall explore the consequences of a model assumption including higher
order terms. To this end we add terms of third and fourth order to the
r.h.s. of equ. \eqref{weak_coupling_T_g}

\begin{equation}
\partial _{\Lambda }g=-2g^{2}(1-\alpha g-\beta g^{2})\tanh \left( \frac{%
\Lambda }{2T}\right) \frac{\Lambda }{\Lambda ^{2}+\Gamma ^{2}}\,,
\label{RG_strong_coupling}
\end{equation}%
with coefficients $\alpha ,\beta $ to be determined by requiring that the
unitarity limit be satisfied and that the resulting conductance fits the
corresponding NRG result \cite{Costi} best. We find that taking $\alpha =0$
and putting $\beta =g_{u}^{2}=4/(3\pi )$ , in equ. \eqref{RG_strong_coupling}
leads to the best agreement with the NRG data on the linear response
conductance \cite{Costi}. Note that the additional terms are not interpreted
as two-loop and three-loop contributions to the $\beta -$function. Rather,
they are thought to represent the zero of the $\beta -$function at the
strong coupling fixed point. A complete discussion of the $\beta -$function
at equilibrium for the anisotropic Kondo model in the strong coupling regime
has been given in \cite{Kehrein2} , on the basis of the flow equation
method. The simple ansatz used here cannot be compared with the entirely
different approach of \cite{Kehrein2}. The point we want to make here is
that a simple adjustment of the $\beta -$function for the coupling constant
is not sufficient to capture the physics of the relaxation rate (see below).

\bigskip In the following we keep the frequency dependence of $%
\Gamma(\omega) $ . \ Within the pseudo fermion projection scheme using the
limit $\lambda \rightarrow\infty$ the self energy acquires a highly
asymmetric form, which causes problems in the numerical solution. We
therefore choose an alternative projection with $\lambda=0,$which requires
to introduce normalization factors accounting for the difference in the
impurity partition function with and without projection. In the calculation
of $\Gamma(\omega)$ in the absence of a magnetic field these factors are not
necessary. This choice has the advantage of particle-hole symmetry, i.e. $%
\Gamma(\omega)=\Gamma(-\omega)$.

The FRG equations (cf. \ref{approx_leading_component}, \ref%
{approx_gamma_tilde} and \ref{approx_Gamma}) then take the following form 
%\begin{widetext}
\begin{eqnarray}
\partial_{\Lambda}g &=& -g^2\left[1-(g/g_u)^2\right] \frac{\Lambda}{%
\Lambda^2+\Gamma^2(\Lambda)}\sum_{\alpha} \tanh\left(\frac{%
\Lambda-\mu_{\alpha}}{2T}\right) \\
\partial_{\Lambda}\Gamma(\omega) &=&\frac{1}{2} \sum_{\epsilon=\pm\Lambda}%
\sum_{\alpha}\left[f_{\alpha}(\epsilon)\tilde{g}^>(\epsilon,\omega)
+(1-f_{\alpha}(\epsilon))\tilde{g}^<(\epsilon,\omega)\right] \\
\partial_{\Lambda} \tilde{g}^>(\epsilon,\omega) &=& \frac{3}{4}g^2 \sum_{%
\tilde{\epsilon=\pm\Lambda}}\sum_{\alpha} \frac{\Gamma(\omega+\epsilon-%
\tilde{\epsilon})}{(\omega+\epsilon-\tilde{\epsilon})^2
+\Gamma^2(\omega+\epsilon-\tilde{\epsilon})}\left(1-f_{\alpha}(\tilde{%
\epsilon})\right) \\
\partial_{\Lambda} \tilde{g}^<(\epsilon,\omega) &=& \frac{3}{4}g^2 \sum_{%
\tilde{\epsilon=\pm\Lambda}}\sum_{\alpha} \frac{\Gamma(\omega+\epsilon-%
\tilde{\epsilon})}{(\omega+\epsilon-\tilde{\epsilon})^2
+\Gamma^2(\omega+\epsilon-\tilde{\epsilon})}\,f_{\alpha}(\tilde{\epsilon})\,
.
\end{eqnarray}
%\end{widetext}

\begin{figure}[tbp]
\vspace{-3mm} 
\begin{minipage}{65mm}
\includegraphics[width=\linewidth]{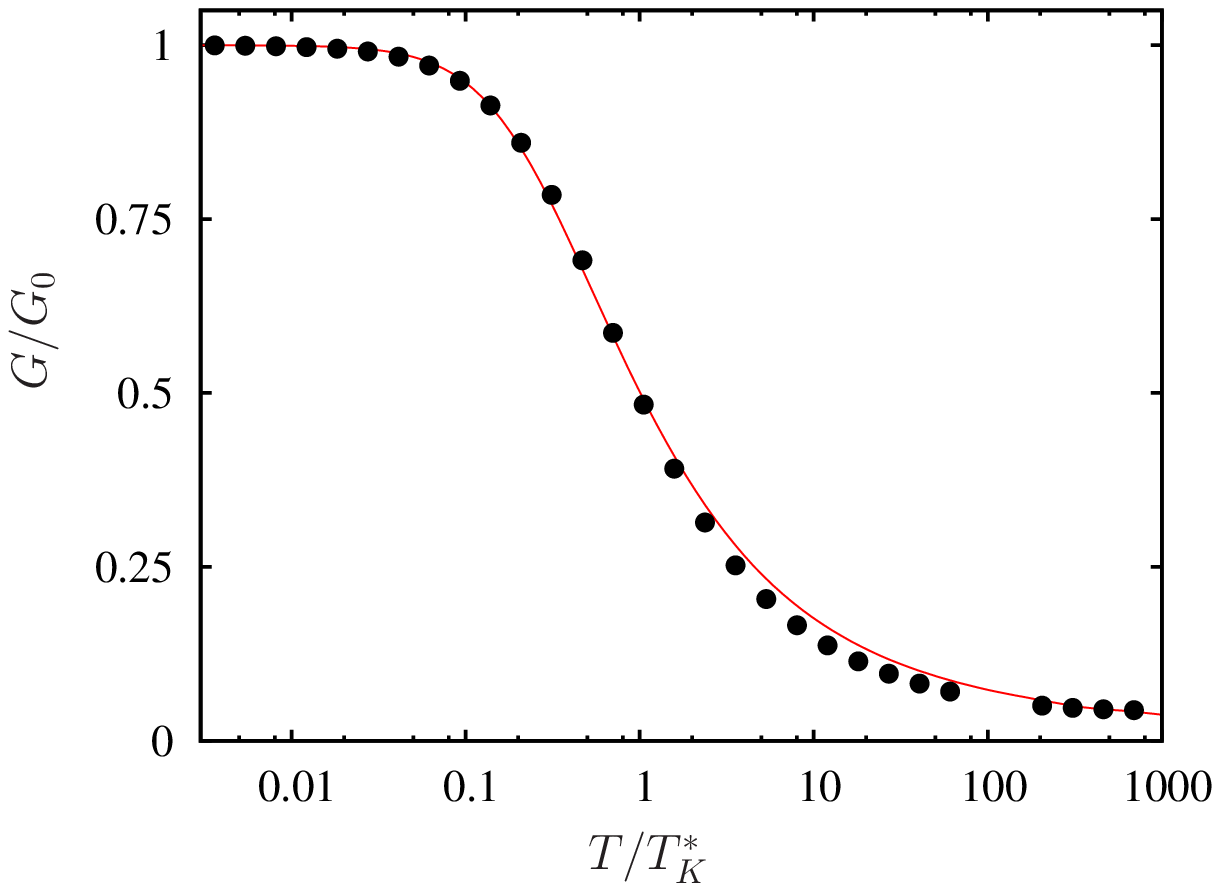}\caption{Comparison
of the linear conductance $G(T)$ from FRG (dots) with the result obtained via
the NRG \cite{Costi}.}
\label{cond_frg_nrg}
\end{minipage}
\hfil
\begin{minipage}{65mm}
\includegraphics[width=\linewidth]{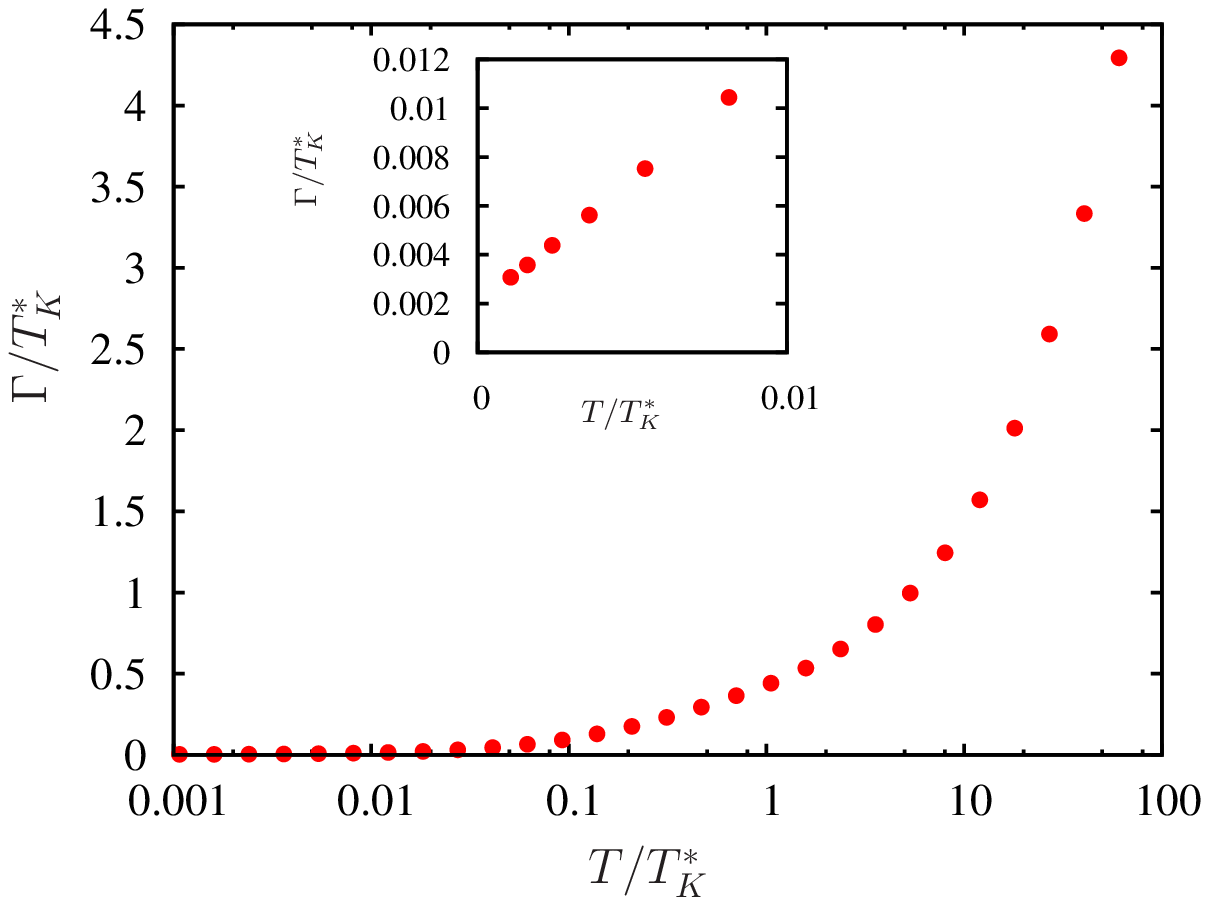}\caption{Imaginary
part of the pseudo fermion self energy $\Gamma(\omega=0)$ versus of $T/T_{K}^{\ast}$
(note the logarithmic scale). Inset: $\Gamma$ at $T/T_{K}^{\ast}\ll 1$ on a linear
scale.}
\label{Gamma_T_all}
\end{minipage}
\end{figure}

Fig. \ref{cond_frg_nrg} shows a comparison of the linear conductance from
FRG (dots) with the NRG data \cite{Costi}. For the purpose of this
comparison we adopted the definition of a Kondo temperature $T_{K}^{\ast}$
used in \cite{Costi} : $G(T_{K}^{\ast})=G_{0}/2$. This differs from our
previous definition of $T_{K}=D_{0}\exp(-1/(2g_{0}))$ by a factor $%
T_{K}^{\ast}/T_{K}\approx2.76$. We find excellent agreement between the FRG
and NRG results. In Fig. \ref{Gamma_T_all} we show the imaginary part of the
pseudo fermion self energy $\Gamma(\omega=0)$ as a function of $%
T/T_{K}^{\ast}$. The inset shows the region $T\ll T_{K}^{\ast}$, where $%
\Gamma$ again decreases almost linearly with decreasing temperature, down to
a similar residual value of $\Gamma /T_{K}^{\ast}\approx0.002$ as found from
the weak coupling FRG equation above.

\begin{figure}[tbp]
\begin{minipage}{65mm}
\includegraphics[width=\linewidth]{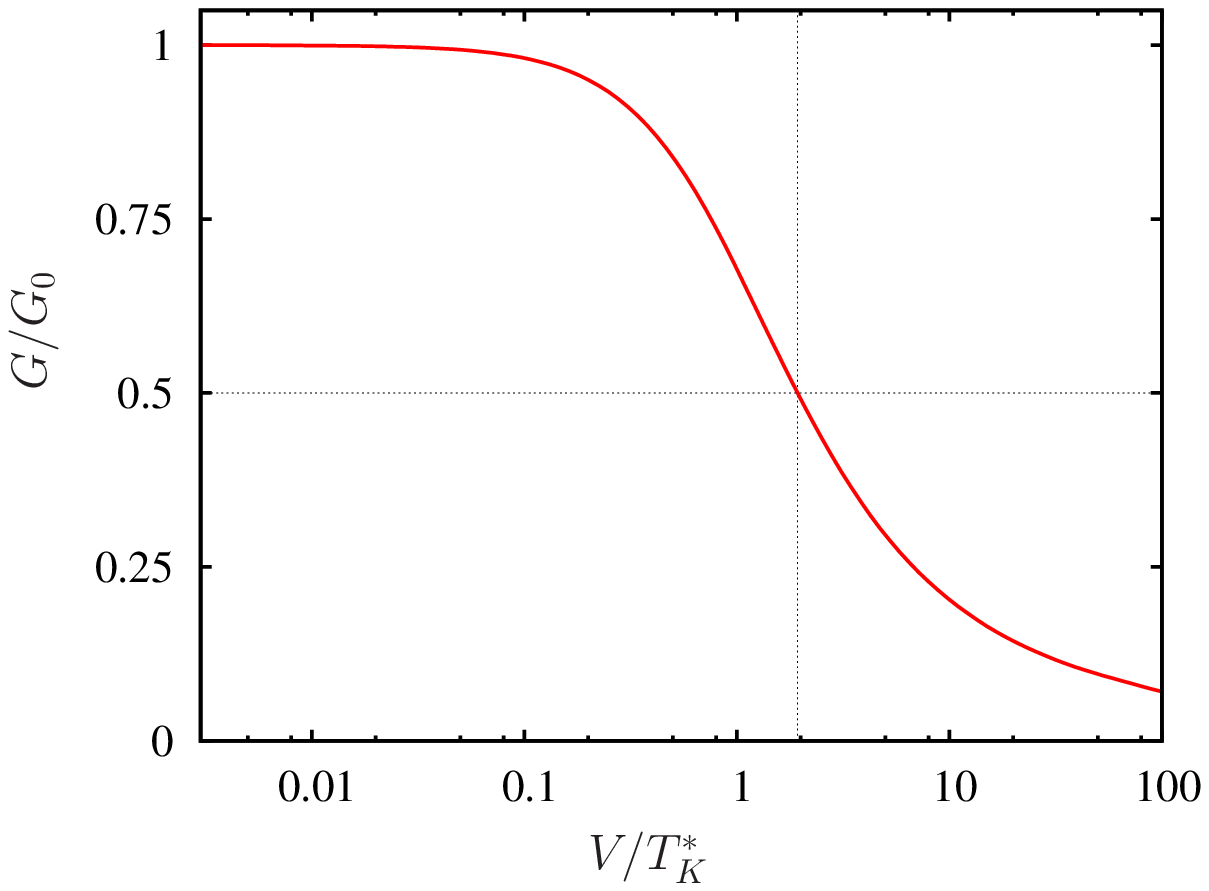}\caption{Conductance
$G(V)$ at $T=0$ from FRG. The lines indicate the half width at half maximum
which amounts to $\Delta V=1.93T_{K}^{\ast}$. }%
\label{cond_frg_V}%
\end{minipage}
\hfil
\begin{minipage}{65mm}
\includegraphics[width=\linewidth]{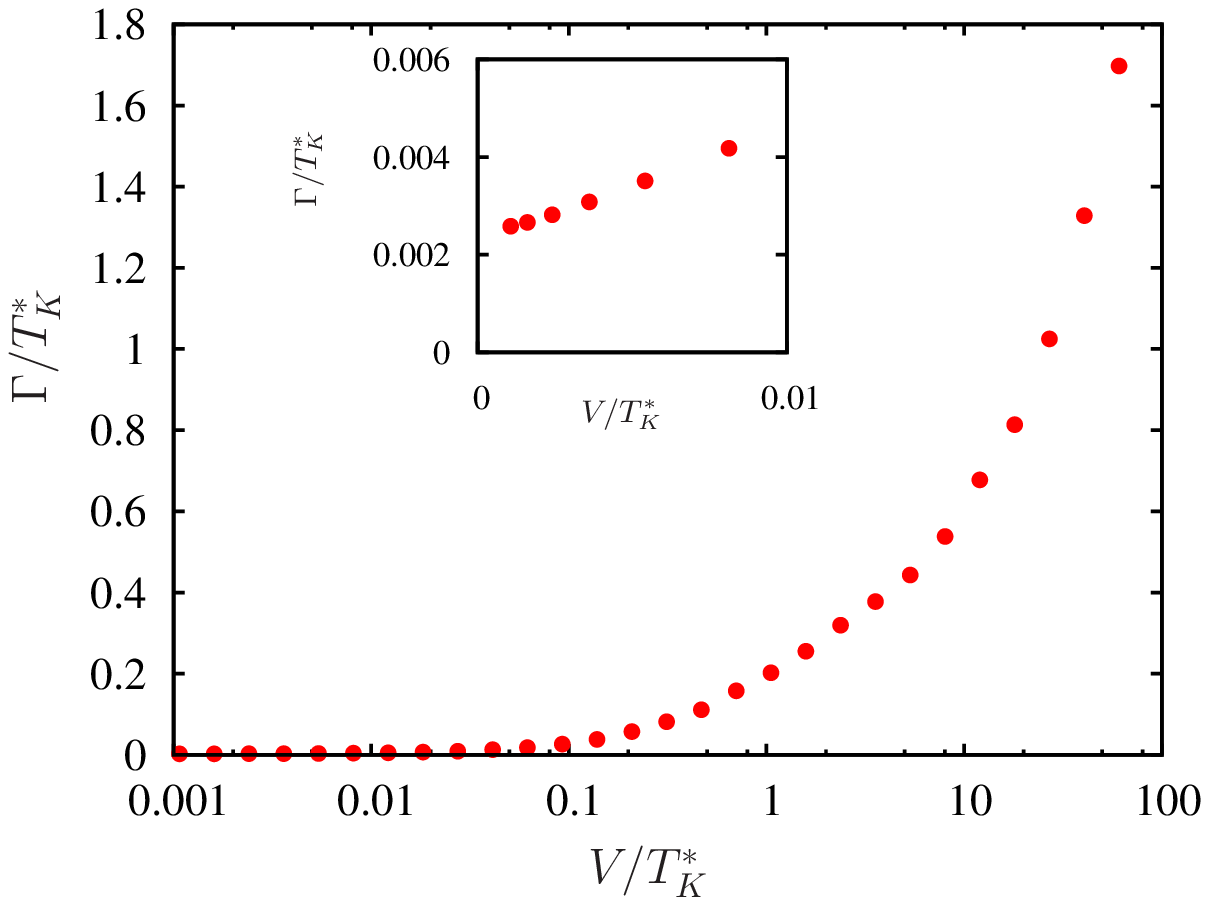}\caption{Imaginary
part of the pseudo fermion self energy $\Gamma(\omega)=0$ versus $V/T_{K}^{\ast}$ on
a logarithmic scale. Inset: $\Gamma(0)$ at $V/T_{K}^{\ast}\ll 1$ on a linear scale. }%
\label{Gamma_V_all}%
\end{minipage}
\end{figure}

We now turn to the nonequilibrium situation: In Fig. \ref{cond_frg_V} we
show the conductance as a function of the voltage. The lines indicate the
half width at half maximum which amounts to $\Delta V=1.93T_{K}^{\ast}$.
Fig. \ref{Gamma_V_all} shows $\Gamma(\omega=0)$ as a function of $%
V/T_{K}^{\ast}$. In the inset $\Gamma$ is seen to drop linearly with
decreasing voltage at \ $V\ll T_{K}^{\ast}$, down to a residual value of $%
\Gamma/T_{K}^{\ast}\approx0.002$.

To summarize, adding a fourth order term with adjusted coefficient to the $%
\beta$-function of the RG equation for the coupling $g(\omega=0)$ leads to
good agreement with the exactly known conductance in the limit $V\rightarrow
0$. The relaxation rate $\Gamma(\omega=0)$, however, comes out three orders
of magnitude too small. We conclude that higher order terms in the $\beta $
-function are not sufficient to cure the problem with $\Gamma$ . In
addition, the frequency dependence of both, $g$ and $\Gamma$ is essential in
reaching the Fermi liquid regime, as we now sketch.

First we note that unitarity does not require $g(\omega)$ to be bounded at
all frequencies. It appears possible to have $g(\omega=0)$ growing large or
even diverging at strong coupling, while the conductance $G$ remains finite.
This is because $G$ as given by equ. \eqref{current} is given by integrating
the product of two $g$ 's over energy. The energy integration removes any
singular behavior of $g$ at $\omega =0$ provided the singularity is not too
strong.

Secondly, a sufficiently strong relaxation rate $\Gamma $ might suppress the
unphysical growth of $g$ . As discussed in detail in section 3.3 the RG
equations in the approximation of taking all energies on shell do not
generate a sufficiently strong $\Gamma (\omega =0)$ at the scale $\Lambda
\simeq T_{K}$ , where it would be needed to cut off the growth of $g$ , for
the reasons discussed above. On the other hand, $\Gamma $ at $\Lambda =0$
turned out to grow too large at $T\ll T_{K}$ . This deficient behavior may
also be remedied by taking the energy dependence of $\Gamma $ and $g$ into
account. Qualitatively one expects that the smearing of the sharp singular
structures in $g(\omega )$ and $\Gamma (\omega )$ by integrations over $%
\omega $ within the width $\Gamma $ specified by the pseudofermion spectral
functions will be equivalent to replacing $g(\omega =0)$ and $\Gamma (\omega
=0)$ by $g(\omega \simeq \Gamma )$ \ and $\Gamma (\omega \simeq \Gamma )$ ,
which are both finite quantities of $O(1)$ and \ $O(T_{K})$, respectively,
not only at the end of the RG process (at $\Lambda =0$ ), but already at $%
\Lambda \approx T_{K}$ . %
%
%
%%%%%%%%%%%%%%%%%%%%%%%%%%%%%%%%%%%%%%%%%%%%%%%%%%%%%%%%%%%%%%%%%%%%%%%%%%%%%%
%%%%%%%%%%%%%%%%%%%%%%%%%%%%%%%%%%%%%%%%%%%%%%%%%%%%%%%%%%%%%%%%%%%%%%%%%%%%%%%
%%%%%%%%%%%%%%%%%%%%%%%%%%%%%%%%%%%%%%%%%%%%%%%%%%%%%%%%%%%%%%%%%%%%%%%%%%%%%%%
%
%
%

\section{Conclusion and outlook}

With the advent of nanophysics in the mid 1990s it became possible to
fabricate nanostructures with the properties of single atoms. In contrast to
real atoms these artificial atoms may be contacted by metallic leads,
opening a new way of probing their properties in transport measurements.
This development led to a revival of Kondo physics, a subject that had
emerged in studies of dilute magnetic alloys. Even though the Kondo effect
had been well understood by the mid 1970s, the new development required
theories describing the Kondo effect out of equilibrium. Most of the
theoretical methods created to describe the Kondo effect in equilibrium do
not work or work only in a limited way for systems out of equilibrium. A
good starting point is perturbation theory in the Keldysh formulation.
However, due to the infrared singular behavior of the Kondo problem, simple
perturbation theory is not sufficient. In equilibrium the powerful method of
the renormalization group has been proposed and successfully implemented
first by Anderson in the perturbative regime and later by Wilson, all the
way to strong coupling.

Anderson's perturbative treatment has been generalized to non-equilibrium in
\ \cite{pertRG1,pertRG2}\ , after recognizing two important new features.
First, in nonequilibrium the coupling constant has to be generalized to an
energy dependent coupling function. Second, the processes limiting the
quantum coherence of the allimportant spin flip processes that are
responsible for the Kondo effect in the first place, become more important
out of equilibrium and may stop the RG flow to strong coupling inside the
weak coupling regime. The latter lead to a decoherence rate, which in \cite%
{pertRG1,pertRG2}\ \ was added ad hoc to the RG equations. In the present
paper the decoherence rate is treated from the beginning on equal footing
with the coupling function. It is known \cite{Paaske2} that the decoherence
rate is given by the physical spin relaxation rate. In technical terms, the
spin relaxation rate has contributions from the self energy as well as from
vertex corrections. Within the standard quantum field theoretical formalism
used in this paper, it is straightforward to incorporate the self energy
contribution. To include the vertex corrections would require to take into
account three particle processes as well. The latter are beyond the scope of
what has been done in the present paper. There exist different formulations
based on the Liouville operator formalism \cite{Schoeller} or the flow
equation method \cite{Kehrein}, which allow to take into account vertex
corrections in a systematic way. However, these methods have their own
drawbacks. The main achievement of the present paper is therefore the
principal demonstration of the appearance and the actual calculation of the
effect of the decoherence rate on the RG flow for the Kondo model.

It is tempting to explore the importance of the decoherence processes in
cutting off the RG flow to strong coupling in equilibrium, where it is
strongest. From the known behavior of the spin relaxation rate as a function
of temperature one would expect that at temperatures at and below the Kondo
temperature the RG flow is already cut off and the coupling constant would
tend to a temperature independent value with $T^{2}$ -corrections at low $T$
. We found something different: apparently the coupling function $%
g(\omega=0) $ grows to values much larger than unity, without, however,
violating the unitarity limit. A careful solution of the frequency dependent
RG equations is a subject for future work. If the coupling grows to large
values one should expect higher order processes to become important. We have
added higher order terms to the RG equation, in such a way that $g(\omega=0)$
tends to a finite value of order unity, recovering the unitarity limit of
the conductance in its simplest form. The resulting conductance curve is in
good agreement with NRG results known for the equilibrium case, and allows
to explore the behavior at finite bias voltage.

\section{Acknowledgement}

We are grateful to Hans Kroha and Achim Rosch for useful discussions in the
early stages of this work. Special thanks are owed to Jens Paaske for
discussions on many aspects of RG in the Kondo problem.

\end{document}